\pdfoutput=1
\documentclass{article}
\usepackage{geometry}
\usepackage[T1]{fontenc}

\usepackage{myoptions}
\usepackage{diagbox}

\usepackage{amsmath,amssymb,amsfonts,mathtools}
\usepackage{etoolbox}
\usepackage{array,booktabs}
\usepackage{microtype}
\usepackage{xcolor}
\usepackage{relsize}
\usepackage{hyperref}
\hypersetup{linktoc = all}
\hypersetup{pdfborderstyle={/S/U/W 0.5}}
\hypersetup{linkbordercolor = gray}
\hypersetup{bookmarksnumbered}
\numberwithin{equation}{section}

\newcommand{\nop}[1]{\mathop{{:}{#1}{:}}\nolimits}

\DeclarePairedDelimiterX\braket[3]{\langle}{\rangle}{#1\mathclose{}\delimsize\vert\mathopen{}\ifblank{#2}{}{#2\mathclose{}\delimsize\vert\mathopen{}}#3}
\DeclarePairedDelimiter{\bra}{\langle}{\rvert}
\DeclarePairedDelimiter{\ket}{\lvert}{\rangle}

\newcommand{\del}{\partial}
\newcommand{\Ocal}{{\mathcal{O}}}

\newcommand{\z}{z}

\newcommand{\zbar}{{\bar{z}}}

\newcommand{\Tbar}{{\bar{T}}}

\newcommand{\Thetabar}{{\bar{\Theta}}}

\newcommand{\Jbar}{{\bar{J}}}

\newcommand{\rmx}{x}

\newcommand{\rmt}{t}

\newcommand{\Hcal}{{\mathcal{H}}}

\newcommand{\TTbar}{\texorpdfstring{\ensuremath{T\Tbar{}}}{TTbar}}

\author{Bruno Le Floch\thanks{\texttt{lefloch@lpt.ens.fr}\quad Philippe Meyer Institute, Physics Department, \'Ecole Normale Sup\'erieure, PSL Research University, 24 rue Lhomond, F-75231 Paris Cedex 05, France} \and M\'ark Mezei\thanks{\texttt{mmezei@scgp.stonybrook.edu}\quad Simons Center for Geometry and Physics, SUNY, Stony Brook, NY 11794, USA}}
\title{Solving a family of \TTbar-like  theories}
\hypersetup
  {
    pdftitle = Solving a family of TTbar-like theories ,
    pdfauthor = Bruno Le Floch and M\'ark Mezei
  }
\begin{document}
\bibliographystyle{jhep}
\maketitle

\begin{abstract}
  We deform two-dimensional quantum field theories by antisymmetric combinations of their conserved currents that generalize Smirnov and Zamolodchikov's $T\bar{T}$ deformation.
  We obtain that energy levels  on a circle obey a transport equation analogous to the Burgers equation found in the $T\bar{T}$ case.
  This equation relates charges at any value of the deformation parameter to
  charges in the presence of a (generalized) Wilson line.
  We determine the initial data and solve the transport equations for antisymmetric combinations of  flavor symmetry currents and the stress tensor starting from conformal field theories. Among the theories we solve is a conformal field theory deformed by $J\bar{T}$ and $T\bar{T}$ simultaneously. We check our answer against results from AdS/CFT.
\end{abstract}

\setcounter{tocdepth}{2}
\tableofcontents

\section{Introduction}

Two-dimensional field theories are interesting theoretical laboratories for discovering new phenomena in quantum field theories. An exciting recent development indicates that in special situations we may control a theory flowing against the renormalization group flow: we can deform a theory by special irrelevant operators and flow towards the ultraviolet without encountering an infinite set of ambiguities that usually plague such attempts. On top of this, in some cases the resulting theory is solvable in the sense that its spectrum on $S^1\times R$ can be determined explicitly in terms of the spectrum of the undeformed theory. This has so far been achieved for the $T\Tbar$ and $J\Tbar$ deformed theories. In this paper, we extend these results to a large family of deformations. Below we briefly summarize what has been understood about these theories in the literature. These exciting findings provide ample motivation for this study.

It was understood in \cite{Zamolodchikov:2004ce} that the composite operator $T\Tbar$ is unambiguously defined in any translationally invariant field theory, because the collision limit of the point splitted operators is regular (up to derivatives). The derivation was extended in \cite{Smirnov:2016lqw} to other operators, and  deforming by such irrelevant operators was proposed. The spectrum of the theory was shown to obey the Burgers equation in \cite{Smirnov:2016lqw,Cavaglia:2016oda}, see also \cite{Cardy:2018jho} for the nonrelativistic case. The spectrum of the $J\Tbar$ deformed theory was obtained in \cite{Chakraborty:2018vja}, see also \cite{Guica:2017lia,Nakayama:2018ujt}. 

One can also arrive at the $T\Tbar$ deformed theories from the point of view of S-matrices. This was developed in \cite{Dubovsky:2012wk,Dubovsky:2013ira,Cooper:2013ffa} and a realization of it as a theory of quantum gravity was proposed in \cite{Dubovsky:2017cnj,Dubovsky:2018bmo,Chen:2018keo}. Other work analyzing the very interesting UV behavior of the theory depending on the sign of the coupling includes \cite{McGough:2016lol,Giveon:2017nie}.

The torus partition function of both the $T\Tbar$ and $J\Tbar$ obeys interesting differential equations, has nice modular properties, and is unique in the appropriate sense \cite{Cardy:2018sdv,Datta:2018thy,Aharony:2018bad,Aharony:2018ics,Dubovsky:2018bmo}. Correlation functions were analyzed in \cite{Cardy:2015xaa,Giribet:2017imm,Kraus:2018xrn,Aharony:2018vux,Guica:2019vnb}.

The first holographic interpretation of  $T\Tbar$ as cutoff AdS$_3$ geometry was proposed in \cite{McGough:2016lol}, progress in this direction is reported in \cite{Shyam:2017znq,Kraus:2018xrn,Cottrell:2018skz}. The holographic interpretation of $J\Tbar$ deformation was studied in \cite{Guica:2017lia,Bzowski:2018pcy}. Higher dimensional generalizations in the holographic context were discussed in \cite{Taylor:2018xcy,Hartman:2018tkw,Shyam:2018sro,Caputa:2019pam}. These ideas were applied to the dS/dS correspondence in \cite{Gorbenko:2018oov}. A second holographic interpretation of a single trace version of $T\Tbar$ with a different sign for the coupling was proposed to describe AdS$_3$ embedded in a linear dilaton background in \cite{Giveon:2017nie}. Soon this approach was generalized to the single trace version of $J\Tbar$ in \cite{Chakraborty:2018vja,Apolo:2018qpq}. Work in this direction includes \cite{Giveon:2017myj,Asrat:2017tzd,Giribet:2017imm,Baggio:2018gct,Babaro:2018cmq,Chakraborty:2018aji,Araujo:2018rho}. The entanglement entropy of these holographic theories was analyzed in \cite{Chakraborty:2018kpr,Donnelly:2018bef}.

 Deformations of supersymmetric theories were discussed in \cite{Baggio:2018rpv,Chang:2018dge}. $T\Tbar$ deformed theories on $S^2$ were analyzed in \cite{Shyam:2018sro,Santilli:2018xux}.   Classical field theories deformed by $T\Tbar$ have interesting properties on their own, which were analyzed in \cite{Cavaglia:2016oda,Bonelli:2018kik,Conti:2018jho,Chen:2018eqk,Conti:2018tca}.

\section*{Organization}

Since our arguments borrow results from a variety of sources, we perform a number of checks and make several comments in the process of solving the spectrum of the deformed theories. We include these at each key step in the paper. To arrive at the result fastest, the reader may wish to follow the argument narrowly and skip the checks and comments, and read Section~\ref{sec:strategy} for the strategy of our approach, Section~\ref{sec:lindef} except for Section~\ref{sec:KdV} for the solution of the theory deformed by linear operators, Section~\ref{sec:GenericPoint} for the universal flow equation describing a generic point in theory space, Section~\ref{sec:Spectrum} except for Section~\ref{sec:String} for the solution of the spectrum, and Section~\ref{sec:Conclusions} for conclusions and future directions.

The content of the rest of the paper is as follows. Section~\ref{sec:KdV} includes our unsuccessful attempt to understand deformations by higher spin (KdV) currents, Section~\ref{sec:checks} contains two complementary checks of the universal equation, and  Section~\ref{sec:String} checks a special case of the spectrum from string theory. The Appendices contain details of our conventions, the worked out example of the compact free scalar, and comparison with the $J\Tbar$ literature.
\\
\\*
{\bf Note added:}  Results equivalent to those in Section~\ref{sec:String} have been obtained independently using the same methods in ongoing work \cite{ongoing}. We thank the authors for comparing our formulas. A summary of that work appears in a coordinated submission to the arXiv \cite{note}. 

\section{A strategy for solving \TTbar-like theories}\label{sec:strategy}

Let us take a 2d QFT on a cylinder, $S^1\times R$, which is translationally invariant in both  the spatial $(S^1)$ and the time $(\R)$ directions. Note that we do not require Lorentz symmetry. In \cite{Zamolodchikov:2004ce,Smirnov:2016lqw} it was shown that there exist composite operators built from conserved currents in any such QFT, whose expectation value factorizes in an energy eigenstate $\ket{n}$:
\es{factor}{
\braket{n}{\ep^{\mu\nu}J^{(1)}_\mu(y) J^{(2)}_\nu(y)}{n}&\equiv\braket{n}{\lim_{x\to y}\ep^{\mu\nu}J^{(1)}_\mu(x)J^{(2)}_\nu(y)}{n}\\
&=\ep^{\mu\nu}\braket{n}{J^{(1)}_\mu}{n}\braket{n}{J^{(2)}_\nu}{n}\,,
}
where in the second line we deleted the arguments to emphasize that the one point functions in energy eigenstates do not depend on the position of the operator. There are two familiar examples of these composite operators: taking $J^{(1)}_z\equiv J$  and $J^{(2)}_{\bar z}\equiv \bar J$ in a CFT, the composite operator $\ep^{\mu\nu}J^{(1)}_\mu J^{(2)}_\nu$ is the exactly marginal operator $J\bar J$, while taking  $J^{(1)}_\mu\equiv T_{1\mu }$ and $J^{(2)}_\mu\equiv T_{2\mu}$ in any 2d QFT,  the composite operator becomes  what is known as \TTbar\ in the literature. (In our conventions, it is $-{1\ov 2\pi^2}T\Tbar$.)

The composite operator $\sO\equiv\ep^{\mu\nu}J^{(1)}_\mu J^{(2)}_\nu$ hence defined can be used to define a one parameter family of theories,
\es{DeformedTheory}{
{d\ov d\lam}S(\lam)=\int d^2x \ \sO_\lam(x)\,,
}
where the notation $\sO_\lam(x)$ serves as a reminder that the conserved currents $J^{(1,2)}_\mu$ building $\sO$ change as $\lam$ is changed. Using factorization, it immediately follows that the energy spectrum of this family of theories obeys
\es{FlowEq}{
{\p\ov \p\lam}E_n=L\,\ep^{\mu\nu}\braket{n}{J^{(1)}_\mu}{n}\braket{n}{J^{(2)}_\nu}{n}\,,
}
where we used the Hellmann-Feynman theorem ${\p\ov \p\lam}E_n=\braket{n}{{\p\ov \p\lam} H}{n}$.
If we want to use this equation, we need to know the matrix elements $\braket{n}{J^{(1,2)}_\mu}{n}$. For the time component $\mu=2$, we have $\braket{n}{J_2}{n}=Q_n/L$, where $L$ is the length of the spatial $S^1$ and $Q$ is the charge corresponding to the conserved current. If $Q$ is the charge of an internal symmetry, its value is quantized, and cannot depend on $\lam$. This includes the case of the momentum along the spatial $S^1$, for which $Q_n=iP_n={2\pi i j_n\ov L}$, where $j_n\in \Z$. Another case is $Q_n=-E_n$. $Q$ can also be a higher spin (KdV) charge of a 2d CFT or integrable model, and to get a closed set of equations, we also need to write down a flow equation for ${\p\ov \p\lam}Q^\text{(higher spin)}_n$ in this case. We treat one such case in a separate publication. 

For the spatial component $\mu=1$ giving $\braket{n}{J_1}{n}$, we do not in general have a physical interpretation. The case of the \TTbar\ deformation of a relativistic field theory is an exception, where we know the value of all matrix elements:
\es{MtxElements}{
\braket{n}{T_{tt}}{n}&=-{E_n\ov L}\,, \qquad \braket{n}{T_{xx}}{n}=-{\p_L E_n}\,, \qquad \braket{n}{T_{xt}}{n}={iP_n\ov L}=\braket{n}{T_{tx}}{n}\,,
}
where the last equality follows from the fact that the stress tensor is symmetric. Plugging these into \eqref{FlowEq} we obtain the Burgers equation of \cite{Smirnov:2016lqw}:
\es{Burgers}{
{\p\ov \p\lam}E_n=\frac12\le(E_n\p_L E_n+{P_n^2\ov L}\ri)\,,
}
where the overall factor on the RHS follows from our choice of normalization of the composite operator $\sO$, as discussed below \eqref{FlowEq}.

We propose to proceed in the more general case, where general considerations do not determine $\braket{n}{J_1}{n}$, by coupling the current to an infinitesimal constant background field
\es{DeformedTheory2}{
\de_a S(\lam,a)\equiv\int d^2x \ i J_1(x)\,.
}
With the introduction of $a$, \eqref{FlowEq} becomes:
\es{FlowEq2}{
{\p\ov \p\lam}E_n={1\ov L}\le({Q^{(2)}}{\de_{a^{(1)}}}E_n-{Q^{(1)}}{\de_{a^{(2)}}}E_n\ri)\,.
}
We do not want to introduce background fields for quantities that we know from other considerations, hence in such cases  it is understood that $\de_{a}E_n$ should be replaced by the appropriate quantity in this equation, see e.g. \eqref{MtxElements}.

\begin{figure}[!ht]
\begin{center}
\includegraphics[scale=0.6]{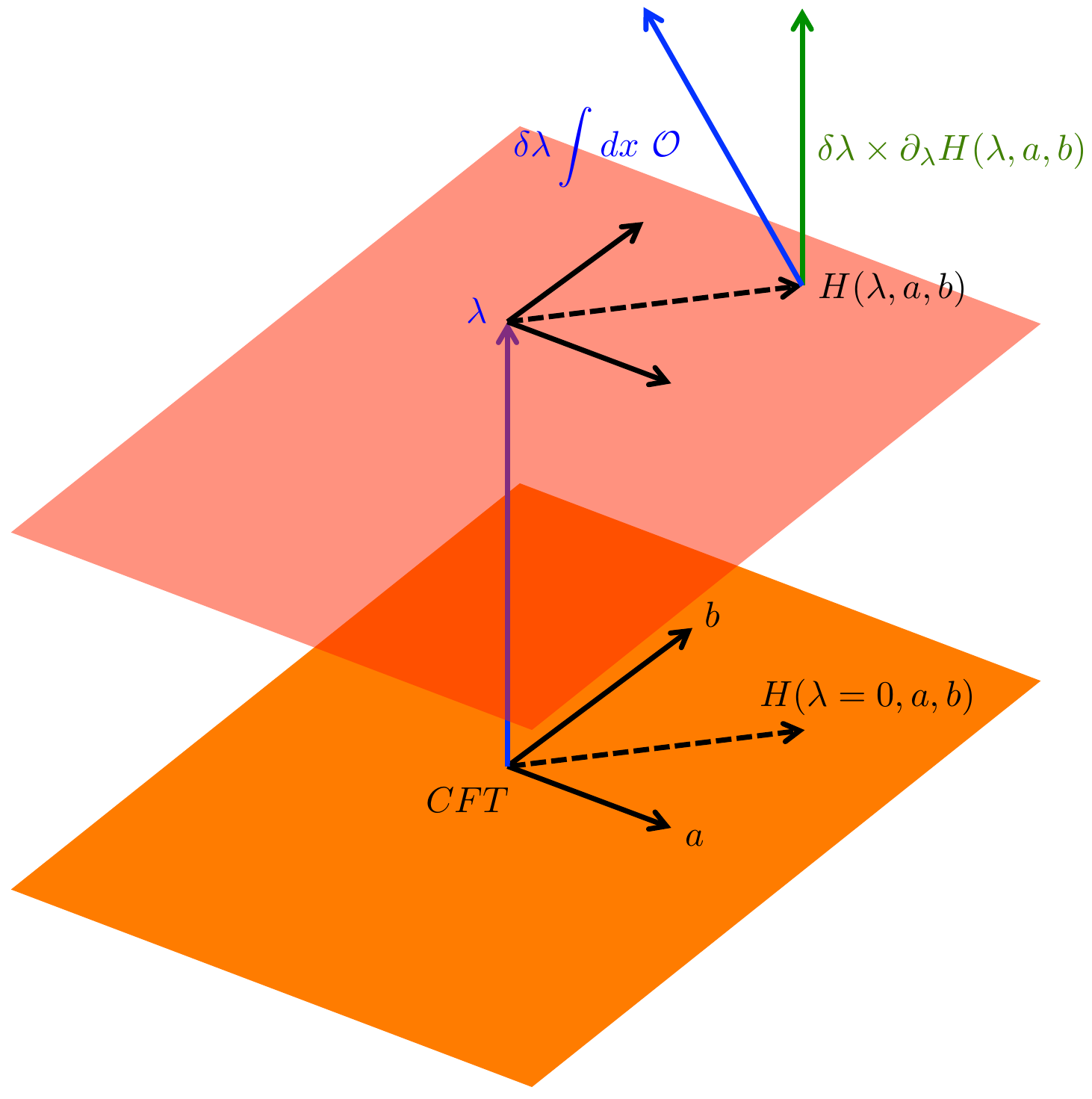}
\caption{Graphical representation of the strategy solving deformations of CFTs by quadratic composite operators. Turning on the background fields $a,b$ determines the initial value surface, drawn here as a bright orange plane. These are the directions corresponding to linear deformations. The $\lam$ direction in coupling space represents the deformation by the quadratic composite operator $\sO$. We erect a coordinate system by first deforming by $\int dx \ \sO_\lam$ as in \eqref{DeformedTheory} and going $\lam$ distance. Subsequently we implement the linear deformations. Hence, deforming a generic point in coupling space by $\de \lam \int dx\ \sO$ (indicated by blue arrow) does not in general agree with  $\de \lambda\times \partial_\lambda H (\lambda,a,b)$. \label{fig:Strategy}}
\end{center}
\end{figure}

In order for \eqref{FlowEq2} to be useful, we have to understand two things. First, to use it as an evolution equation, we need to understand the equation away from infinitesimal $a^{(I)}$. In this paper we will often refer to these deformations as linear deformations: what we mean by this is that the deforming operator is linear in the spatial component of the current. We work to all orders in  $a^{(I)}$, not just first (linear) order. We will refer to the deformations by the quadratic composite operators $\sO$ as quadratic deformations; again we work to all orders in $\lam$, not just to second (quadratic) order. The deformations do not in general commute, namely the vector fields describing the flow in coupling space have a nonzero Lie bracket. We want to understand the flow in some coordinate system in coupling space $(\lam, a^{(I)})$ taking into account this noncommutativity. Second, if we want to solve the $\lam$-evolution in this enlarged coupling space, we need to understand the theory not just at $S(\lam=0)$, but at $S(\lam=0,a^{(I)})$. This can be done if the theory at $\lam=a^{(I)}=0$ is a CFT, because for holomorphic (antiholomorphic) currents, $J_1=\mp iJ_2$. Besides all these challenges, we have to make sure that ambiguities (e.g., improvement transformations) do not ruin the universality of the result. In Figure~\ref{fig:Strategy} we give an illustration of our strategy.

In what follows, we present strong arguments that the outlined strategy works for a large family of irrelevant deformations of CFTs. We remark that \cite{Chakraborty:2018vja} solved the $J\Tbar$-deformed theory using a different method: the existence of a holomorphic current. We reproduce their results in our framework. We explain how to reconcile the two viewpoints in Appendix~\ref{app:JTbar}.

\section{Linear deformations}\label{sec:lindef}

In what follows, we find it convenient to work in the Hamiltonian formalism on $S^1\times \R$ with objects understood to be operators. The conservation equation of a current $J_\mu$ in our conventions is:
\es{ConsEq3}{
0=\p_x J_x + [H,J_t]\,.
}
Our conventions are collected in Appendix~\ref{app:conv}.

\subsection{\label{sec:CFT+stress}CFT deformed by stress tensor}

Using translational invariance, we know the diagonal matrix elements of the stress tensor \eqref{MtxElements}, except for that of $T_{tx}\neq T_{xt}$ since we do not assume Lorentz invariance of the deformed theory. According to the general strategy, we introduce a constant background field $b$ for this operator. We want to determine $H(b)$ for finite $b$. Its evolution equation is 
\es{DefSb}{
{\p\ov \p b} H(b)= -i \int dx \ T_{tx}(x)\,.
}
Note that in Euclidean signature $T_{tx}$ is antihermitian, hence the $i$ in the above formula. We were not able to determine $H(b)$ in closed form, if we only assume Lorentz symmetry for $H(b=0)$. If $H(b=0)$ describes a CFT, however, we obtain a solvable system of equations. We only need to assume conformal symmetry at the starting point of the flow equation \eqref{DefSb}. Away from $b=0$ the stress tensor will not be symmetric as can be seen from the explicit expressions we give below.

Starting from a CFT we have $T_{\rmt\rmt}^{(0)}+T_{\rmx\rmx}^{(0)}=0$ (at zero deformation) in addition to $T_{\rmt\rmx}^{(0)}=T_{\rmx\rmt}^{(0)}$.
Using the definition \eqref{DefSb}, we get $\del_bT_{\rmt\rmt}=iT_{tx}$. Because momentum is quantized and hence cannot depend on $b$, we have $\del_bT_{xt}=0$. Using the conservation equations $\del_\rmx T_{\mu\rmx} = - [H,T_{\mu\rmt}]$, we work out $T_{\rmx \rmt}=T_{\rmx \rmt}^{(0)}$, and
\begin{equation}
  \begin{aligned}
    T_{\rmt\rmt} &= - T_{\rmx\rmx}  = \frac{1}{1-b^2} T_{\rmt\rmt}^{(0)} + \frac{ib}{1-b^2} T_{xt}^{(0)} , \\
    T_{\rmt\rmx} & = \frac{-2ib}{(1-b^2)^2} T_{\rmt\rmt}^{(0)} + \frac{1+b^2}{(1-b^2)^2} T_{xt}^{(0)} \,,
  \end{aligned}
\end{equation}
where we only had to use \eqref{ConsQuant}  that gives $H,\, P$ in terms of the components of the stress tensor, and \eqref{Trans} that gives the spacetime translations they generate.
Interestingly $T_{\rmt\rmt}+T_{\rmx\rmx}=0$ for all~$b$. Integrating $T_{tt},\, T_{xt}$ as in \eqref{ConsQuant} we find:
\begin{equation}\label{HunderTdefintermsofP1}
  P = P^{(0)} , \qquad
  H = \frac{H^{(0)}+bP^{(0)}}{1-b^2}\,.
\end{equation}
It will be helpful to rewrite the result for $H$ as:
\es{Ppm1}{
H&=-\le(\frac{1}{1- b} P^{(0)}_{1}+\frac{1}{1+ b} P^{(0)}_{-1}\ri)\,,\\
P^{(0)}_{\pm 1}&\equiv-{H^{(0)}\pm P^{(0)} \ov2}\,,
}
which we interpret to say that the initial value of the holomorphic and antiholomorphic charges contribute to $H$ weighted by the factor $\frac{1}{1\mp b}$.

\subsection{\label{sec:CFT+J+stress}CFT deformed by currents and the stress tensor}

We consider a CFT with left- and right-moving $U(1)$ symmetry currents $J_\mu$ and $\bar J_\mu$. The case of a CFT deformed by only $J_x$ is straightforward, so we move on to discussing a CFT deformed by  $J_x$, $\bar J_x$, and $T_{tx}$. In familiar deformations of CFTs, we are used to losing one of the conserved currents. In contrast, a curious feature of the deformed theories that we consider is that the currents $J_\mu$ and $\bar J_\mu$ remain separately conserved. 

We will see that it is possible to keep the corresponding conserved charges unchanged under the deformation. An easy way to argue for this is to take an example where the charges generate a compact $U(1)\times U(1)$ symmetry: the spectrum of charges cannot depend on the deformation due to charge quantization. Such an example is provided by the compact scalar discussed in Appendix~\ref{app:scalar}. Since our methods do not depend on global aspects, we then expect that the spectrum of charges corresponding to internal symmetries does not change. We will check by explicit computation that this is indeed the case.

  We observe a major simplification during the derivation: the linear deformations commute (to all orders in the background fields $a,\,\bar a,\,b$). We derive this using explicit computation. An explanation for this is that all operators that feature in the derivation are neutral and hence commute with $Q,\, \bar Q$. Thus adding them to the Hamiltonian does not change the conservation equation, and the currents $J_\mu,\, \bar J_\mu$ remain unchanged under the $a,\, \bar a$ deformation. Noncommutativity, however, will be an essential aspect of the physics of the coupling space flow once we include quadratic deformations in Section~\ref{sec:GenericPoint}.
    
We expect based on \eqref{Ppm1} that the conserved charges behave as:
\es{AllCharges}{
Q(a,\bar a,b)&= Q^{(0)}\,, \qquad \bar Q(a,\bar a,b)= \bar Q^{(0)}\,, \qquad P(a,\bar a,b)= P^{(0)}\,,\\
H(a,\bar a,b)&=  \frac{H^{(0)}+bP^{(0)}}{1-b^2}+ {a Q^{(0)}\ov 1-b}+ {\bar a \bar Q^{(0)}\ov 1+b}\,,
}
where we note that the internal symmetry charges $Q^{(0)},\bar  Q^{(0)}$ are pure numbers (see Appendix~\ref{app:scalar} for their allowed values for the example of the compact scalar), while $P^{(0)}\in {2\pi\ov L} \Z$ depends on $L$. It is only the second line that is an Ansatz, the first line follows from general principles. 

A little bit of thought leads to the following Ansatz for the currents that we will verify below:
\es{CurrentGuess}{
T_{\rmt\rmt}&=\frac{1}{1-b^2}T_{\rmt\rmt}^{(0)}+\frac{ib}{1-b^2}T_{xt}^{(0)}-\frac{1}{1-b}a J^{(0)}_t-\frac{1}{1+b}\bar a \bar J^{(0)}_t\,,\\
T_{tx}&=-\frac{2ib}{(1-b^2)^2} T_{\rmt\rmt}^{(0)} + \frac{1+b^2}{(1-b^2)^2} T_{xt}^{(0)}+{i\ov (1-b)^2}a J^{(0)}_t-{i\ov (1+b)^2} \bar a \bar J^{(0)}_t\,, \\
J_x&=-{i\ov 1-b}J^{(0)}_t\,, \qquad \bar J_x={i\ov 1+b}\bar J^{(0)}_t\,.
}
The components $J_t, \bar J_t, T_{xt}$ cannot depend on the background fields because of the quantization conditions \eqref{AllCharges}, while we will not need $T_{xx}$. The Ansatz clearly obeys
\es{AnsatzEq}{
\del_bT_{tt}&=iT_{tx}\\
\del_aT_{tt}&=-iJ_{x}\,, \qquad \del_{\bar a}T_{tt}=i\bar J_{x}\,.
}

Let us verify that the Ansatz indeed solves the problem. The current conservation equation is
\es{CurrCons}{
\p_x J_x+[H,J_t]&=-{i\ov 1-b}\p_x J^{(0)}_t+ \le[\frac{H^{(0)}+bP^{(0)}}{1-b^2}+ {a Q^{(0)}\ov 1-b}+ {\bar a \bar Q^{(0)}\ov 1+b}, J^{(0)}_t \ri]\\
&=-{i\ov 1-b}\p_x J^{(0)}_t+ \le({1\ov1-b^2 }\le[H^{(0)}, J^{(0)}_t \ri]-{ib\ov 1-b^2}\p_x J^{(0)}_t\ri)\\
&={1\ov1-b^2 } \le(\p_x (-iJ^{(0)}_t)+\le[H^{(0)}, J^{(0)}_t \ri]\ri)=0\,,
}
where in the first line we plugged in the Ansatz, in the second we used the fact that $[Q^{(0)},J^{(0)}_t]=[\bar Q^{(0)},J^{(0)}_t]=0$ and that $[P,\sO]=i\p_x \sO$, and in the third we discovered the original conservation equation; recall that $J_x^{(0)}=-iJ^{(0)}_t$. Similarly, recycling the results of the previous section and using that $[Q^{(0)},J^{(0)}_t]=[\bar Q^{(0)},J^{(0)}_t]=0$, we learn that in the stress tensor conservation equation we can focus on the linear in $a$ terms:
\es{CurrCons2}{
&\le(\p_x T_{tx}+[H,T_{tt}]\ri)\big\vert_\text{linear in $a$}\\
&={i\ov (1-b)^2} a \p_x J^{(0)}_t+ \le[ {a Q^{(0)}\ov 1-b}, \frac{1}{1-b^2}T_{\rmt\rmt}^{(0)}+\frac{ib}{1-b^2}T_{xt}^{(0)} \ri]+\le[ \frac{H^{(0)}+bP^{(0)}}{1-b^2},-\frac{1}{1-b}a J^{(0)}_t\ri]\\
&=ia\le({1\ov (1-b)^2}  \p_x J^{(0)}_t+{i\ov (1-b^2)(1-b)}\le[ H^{(0)}, J^{(0)}_t\ri]-{b\ov (1-b^2)(1-b)}\p_x J^{(0)}_t\ri)=0\,,
}
where in the second equality we used that $[Q^{(0)},T_{\mu\nu}^{(0)}]=0$, and in the third that $\le[H^{(0)}, J^{(0)}_t \ri]=i\p_x J^{(0)}_t$.

\subsection{Ambiguities}\label{sec:Ambiguity}

There are ambiguities in the determination of currents from conservation laws. We could perform an improvement transformation on the currents, $J_\mu=J_\mu^\text{(min)}+\ep_{\mu\nu}\p^\nu \chi$ with an arbitrary scalar function $\chi$, which neither violates conservation nor changes the value of the conserved charge. We also note that improvement only changes the quadratic composite operators $\sO=\ep^{\mu\nu}J_\mu^{(1)}J_\mu^{(2)}$ by total derivatives, hence the theories deformed by $\sO$ and $\sO^\text{(min)}$ are equivalent. Another ambiguity arises from mixing two conserved currents.\footnote{In the case when these currents generate a nonabelian symmetry algebra, this mixing can lead to a redefinition of the Cartan. We do not encounter this issue in this paper.} In the absence of Lorentz invariance mixing $J_\mu$ and $T_{\mu\nu}$ is allowed: an example that arises in the discussion of Appendix~\ref{app:JTbar} is the redefinition $\hat J_\mu\equiv J_\mu -2\pi^2 i \ell \, T_{\zbar \mu}$. Finally, we could simply multiply the conserved current by an arbitrary constant $\alpha$ to get a new conserved current: $\hat J_\mu\equiv \alpha J_\mu$.

 We have fixed all these ambiguities above by requiring that not just the charge $Q$, but also the time component of the current $J_t$ remains unchanged. An additional ambiguity that arises is the allowed shift of the spatial component of currents by multiples of the identity. The most general such transformations are:
\es{ShiftsByIdentity}{
T_{tt}&=T_{tt}^\text{(min)}+f_1(b) a^2+2 f_2(b)a \bar a+f_3(b) \bar a^2\,,\\
J_x&=J_x^\text{(min)}+2i(f_1(b) a+f_2(b) \bar a)\,, \qquad \bar J_x=\bar J_x^\text{(min)}-2i(f_2(b) a+f_3(b) \bar a)\,,\\
T_{tx}&=T_{tx}^\text{(min)}-i\le(f_1'(b) a^2+2 f_2'(b)a \bar a+f_3'(b) \bar a^2\ri)\,,\\
T_{xx}&=T_{tx}^\text{(min)}+f_4(b) a^2+2 f_5(b)a \bar a+f_6(b) \bar a^2\,,
}
where $\sO^\text{(min)}$ refers to the expressions given in \eqref{CurrentGuess}. These shifts satisfy dimensional constraints and the defining equations \eqref{AnsatzEq}. We do not know of any algebraic way to fix these ambiguities. Using the Lagrangian formulation, however in  in Section~\ref{sec:AmbiguityFix}, we will fix these ambiguities.

\subsection{An attempt to deform by KdV currents}\label{sec:KdV}

Encouraged by the success with linear deformations by $T_{tx}$ and $J_x$, we attempt to make linear deformations by the higher spin KdV currents. For concreteness, we take the simplest one, obeying the conservation equation
\es{T4}{
\bar \p T_4&= \p \Theta_2\,,
}
which in more convenient coordinates takes the form
\es{T42}{
0&=\p_x J_x^{(4)}+\p_t J_t^{(4)}\,,\\
J^{(4)}_\mu&=\le(J_x^{(4)},J_t^{(4)}\ri) =\le(-2\pi i(T_4-\Theta_2),2\pi(T_4+ \Theta_2 )\ri)\,.
}
We will use that $\p_t J_t^{(4)}=[H,J_t^{(4)}]$ in the canonical formalism.

The corresponding conserved charge is
\es{P_3}{
P_3&=\int dx\ J^{(4)}_t(x)\,.
}
We used the similarly defined conserved charges $P_{\pm1}$ in \eqref{Ppm1}. The KdV conserved charges are mutually commuting, $[P_s,P_\sig]=0$.

We want to introduce a background field $\al$ that couples to $J^{(4)}_x$, i.e.
\es{Hal}{
{\p H(\al)\ov \p \al}=i \int dx \ J^{(4)}_x\,.
}
Specializing the argument of \cite{Smirnov:2016lqw} to this case shows that all the $P_s$ can be preserved under this deformation: First, it is more convenient to work in the path integral formalism and define
\es{PsContour}{
P_s\equiv {1\ov 2\pi}\oint_C \le(dz\ T_{s+1}+d\bar z \ \Theta_{s-1} \ri)\,,
}
and then from $[P_s,P_\sig]=0$ it follows that
\es{Adef}{
[P_s,T_{\sig+1}(z)]=\p A_{s,\sig}(z) \,, \qquad [P_s,\Theta_{\sig-1}(z)]=\bar \p A_{s,\sig}(z)\,.
}
 Second, we assume that the theory at $\al$ has a conserved current $J_\mu^{(4)}$, and ask if we can adjust the current so that it remains conserved at  $\al+\de \al$, and that its charge commutes with the Hamiltonian.
We now work out how $P_3$ changes under this deformation. We write:
\es{CommChange}{
0&=\de [H,P_3]=[\de H, P_3]+[H, \de P_3]\,,\\
[H, \de P_3]&=i\int dx\ [P_3, J^{(4)}_x(x)]=-2\pi i \int dx\  \p_t A_{3,3}(x)=[H, -2\pi i \int dx \  A_{3,3}(x)]\,,
} 
where we used \eqref{T42} and \eqref{Adef}. From this we conclude that up to total derivatives
\es{deJt}{
\de J_t^{(4)}(x) = -2\pi i A_{3,3}(x)\,.
}
With a bit more work, it is possible to determine how to adjust $\de J_x^{(4)}$ by local operators so that that the current remains conserved. 

Now we list some formulas valid at the CFT point. The KdV currents are well known, and $A_{3,3}$ can be computed using the definition \eqref{Adef}:
\es{Ts}{
T_2&=T\, \qquad T_4=\nop{T^2}\,, \qquad T_6=\nop{T^3}+{c+2\ov 12}\nop{(\p T)^2}\,, \qquad \dots\\
A_{3,3}&=-4i \nop{T^3}+{i(c+2)\ov 2}\nop{(\p T)^2}+\text{(tot. der.)}=-4iT_6+{i5(c+2)\ov 6}\nop{(\p T)^2}+\text{(tot. der.)}\,.
}
Using these formulas, for the family of theories defined by \eqref{Hal} to first order in $\al$ we get:
\es{FirstOrder}{
H&=H^{(0)}+2\pi \al P_3^{(0)}+O(\al^2)\,,\\
P_3&=P_3^{(0)}-8\pi \al \le(P_6^{(0)}-{5(c+2)\ov 24}\int dx \nop{(\p T^{(0)})^2} \ri)+O(\al^2)\,,
}
where the $(0)$ superscript indicates CFT quantities. Since $\int dx \nop{(\p T^{(0)})^2} $ does not commute with $P_s^{(0)}$, unlike in the previously considered cases, we cannot use the CFT eigenstates that simultaneously diagonalize $P_s^{(0)}$ as the eigenbasis for the linearly deformed theories.\footnote{We emphasize that $[H,P_3]=0$ to all orders in $\al$, they just do not commute with  $P_s^{(0)}$. To see that $[H,P_3]=0$ to $O(\al^2)$ in \eqref{FirstOrder}, the only nontrivial step involves realizing that  $\le[H^{0}, \int dx \nop{(\p T^{(0)})^2}\ri]=0$, which is true because $\nop{(\p T^{(0)})^2}$ is holomorphic.} This in itself does not constitute a no-go result, as in later sections we will be able to solve for the spectrum of Hamiltonians that do not commute with $H^{(0)}$. However, we were not able to understand how to extend \eqref{FirstOrder} to all orders in $\al$ and how to  obtain the spectrum of \eqref{FirstOrder} efficiently. It would be very interesting to make progress on these fronts.

\section{Understanding the flow around a generic point}\label{sec:GenericPoint}

We saw that the vector fields generating the flow in the directions of the linear deformations commute. This is not true once we include the quadratic composite operators. Let us denote by $\sH(\lam,a,\bar a, b)$ the Hamiltonian density that we obtain by first doing quadratic, {\it and then} linear deformations. See Figure~\ref{fig:Strategy} for a graphical representation. We want to determine  $\p_\lam \sH(\lam,a,\bar a, b)$ and solve the resulting equation using the initial conditions determined in Section~\ref{sec:lindef}. The existence of such a universal equation valid for all theories is already nontrivial, but the equation itself will have even more structure.
Schematically, we will find that for every quadratic operator (and their linear combinations)
\es{KeyEq}{
\p_\lam \sH(\lam,a,\bar a, b)=g_1(b)\cdot \sO_1+  a g_{2}(b)\cdot\sO_{2}+ \bar a  g_{3}(b) \cdot \sO_3+a^2  g_4(b) \cdot  \sO_4+  a \bar a g_5(b)\cdot  \sO_5+  \bar a^2 g_6(b)\cdot  \sO_6\,,
}
where we lightened the notation by introducing $g_I(b)\cdot \sO_I= \sum_{i} g_{Ii}(b) \sO_{Ii}$. This extra sum is necessary, since for different operators $\sO_{Ii}$ the $b$ dependence of the coefficient can be different.  A remarkable property is that the RHS does not depend on $\lam$ explicitly. Since $[\lam]\leq 0$ (the exact value depends on what operator we are deforming by) and  $[a,\bar a]=1, \, [b]=0$, positive powers of $\lam$ would multiply high dimension operators (or high powers of $a,\bar a$) on the RHS. The absence of $\lam$ severely restricts the structure of the RHS: we cannot have too high powers of $a,\bar a$ on dimensional grounds, and in practice $a,\bar a$ only features quadratically. We note that there is another dimensionful quantity in the problem, the length of the spatial $S^1$, $[L]=-1$. Since we are (at least formally) working in local field theory, it cannot appear in \eqref{KeyEq}. In order for \eqref{KeyEq} to be unambiguous, we need the operators $\sO_i$ to be either $J_\mu$, $\bar J_\mu$, $T_{\mu\nu}$ or one of the factorizing quadratic composite operators built from them, as generic composite operators have arbitrariness in their definitions. $\sO_i$ are indeed such special operators. 

We note that if there existed a universal equation for deformations by higher spin KdV currents similar to \eqref{KeyEq}, we expect that it would involve an infinite number of terms on the RHS.  This has simple dimensional reasons. The background fields coupling to the spin-$s$ higher spin currents have dimension $[\al_s]=2-s$, and hence for $s>2$ it is an irrelevant coupling. We expect that arbitrary high powers of it would appear on the RHS with very irrelevant factorizing composite operators built from the KdV currents multiplying them. It would be interesting to understand, if a universal equation exists at all, and whether our expectations about it are realized.

Once we have the operator equation \eqref{KeyEq}, we can take the diagonal matrix element in the eigenstate $\ket{n}$, use the Hellmann-Feynman theorem on the LHS as in \eqref{FlowEq}, factorization on the RHS for composite operators, and compute the matrix elements according to what was explained in Section~\ref{sec:strategy}. This then leads to a flow equation for the energy eigenvalues in the enlarged coupling space.

We are not able to derive \eqref{KeyEq} in a systematic manner. We will find the equation for deformations starting from the classical free scalar theory nonperturbatively in Section~\ref{sec:Classical}. We then check the validity of the equation in a more general classical field theory in Section~\ref{sec:CLcheck}, and  in the quantum theory at low orders in perturbation theory in Section~\ref{sec:PertCheck}. We solve the equations in Section~\ref{sec:Solution}. The solution reproduces the energy spectrum of the $T\Tbar$ and $J\Tbar$ deformed theories obtained previously in the literature as special cases. We also compute the spectrum of a certain string theory in Section~\ref{sec:String}, which was argued in \cite{Chakraborty:2018vja} to be dual to a theory that is closely related to a CFT deformed by $T\Tbar,\, J\Tbar,\, \bar J T$ simultaneously; the results are again in perfect agreement. 

\subsection{Flow equation for the classical free scalar}\label{sec:Classical}

It is clear that if a universal operator equation like \eqref{KeyEq} exists, then it must hold for classical field theories. Conversely, we can use classical field theory to conjecture the equation \eqref{KeyEq}, and then test it in the quantum theory. There also exists a way to read off the energy levels from the knowledge of the classical Hamiltonian, assuming that a universal expression for these also exists, see Appendix~\ref{app:recover}.

To keep the discussion simple, we will study the case of the $J\Tbar$ deformation of the free massless scalar first. To reiterate, we want determine the Hamiltonian density $\Hcal(\lambda,a,b)$ by first deforming by $\lambda J\Tbar$, and then by $a J_\rmx$ and by~$b T_{tx}$. This implies that $\del_\lambda \Hcal(\lambda,a,b)$ is not only $J\Tbar$. As we will see it is instead a linear combination of various deformations, see also Figure~\ref{fig:Strategy}. For our conventions for the free scalar see Appendix~\ref{app:scalar}, this helps explain some signs that appear below.

Now take $\Hcal=h(\del_\rmx\phi,\Pi)$. We enforce the quantization of charges and momentum by requiring that the $t$ components of the corresponding currents do not depend on the couplings $\lam,a,b$ and obtain their $x$ components from conservation:
\es{ConsCurrComps}{
  J_\rmt &= -\frac{1}{2}(\del_\rmx\phi-4\pi \Pi)\,, \qquad
  J_\rmx =2\pi i\biggl(\frac{\del h}{\del(\del_\rmx\phi)} -{1\ov 4\pi} \frac{\del h}{\del \Pi}\biggr) \,,
  \\
  \Jbar_\rmt &= -\frac{1}{2}(\del_\rmx\phi+4\pi \Pi) \,, \qquad
  \Jbar_\rmx = -2\pi i\biggl(\frac{\del h}{\del(\del_\rmx\phi)} +{1\ov 4\pi} \frac{\del h}{\del \Pi}\biggr) \,,
  \\
  T_{\rmt\rmt} &= - h \,, \qquad \qquad\ \,
  T_{tx} = -{i}\frac{\del h}{\del\Pi} \frac{\del h}{\del(\del_\rmx\phi)} \,, \\
  T_{xt}& = - i\,  \Pi \del_\rmx\phi \,, \qquad
  T_{\rmx\rmx} = \Pi \frac{\del h}{\del\Pi} + \del_\rmx\phi \frac{\del h}{\del(\del_\rmx\phi)} - h \,.
}
It is easy to check that the currents are conserved using Hamilton's equations \eqref{HamEoM}, and they reduce to their free scalar counterparts listed in Appendix~\ref{app:scalar}.

Deforming $\Hcal$ by $J_\rmx$ or $\Jbar_\rmx$ amounts to shifting $\del_\rmx\phi$ and $\Pi$.
In particular, $\Hcal(a)=h(\del_\rmx\phi-2\pi a,\Pi+a/2)$ obeys
\es{JxJbarx}{
  \frac{\del\Hcal(a)}{\del a} =  iJ_\rmx (a)\,,
}
where $J_\rmx(a)$ is the spatial component of the current in the presence of the background gauge field~$a$.
The deformation of $\Hcal$ by $ibT_{tx}$ cannot be written in a closed form in general,
\es{bguess}{
  \Hcal(b) = h - b \frac{\del h}{\del(\del_\rmx\phi)} \frac{\del h}{\del\Pi} + O(b^2)\,,
  \quad \text{so that} \quad
  \frac{\del\Hcal(b)}{\del b} =- i T_{tx}(b) \,.
}

After this preparation, consider any deformation with the background fields $a,b$ set to zero:
\es{SDef}{
\del_\lambda \Hcal(\lam) = S\le(\Hcal,{\p \Hcal\ov \del(\del_\rmx\phi)},{\p \Hcal\ov \del\Pi},\del_\rmx\phi,\Pi\ri)
}
 for some $S$ that is a function of its five argument. Note that all the components of the currents in \eqref{ConsCurrComps} are of this form, hence the linear and quadratic deformations of interest in this paper are a special case of $S$. 
Let $\Hcal(\lambda,a,b)$ be obtained by turning on background fields $a$ and~$b$ \emph{after} deforming by $\lambda$. Below, in \eqref{dellambdaHlambdaab}, we give an explicit formula for $\del_\lambda \Hcal(\lambda,a,b)$ in terms of $S$, which is the main result of this section.

Let us define
\es{STilde}{
\tilde S(a,b,\del_\rmx\phi,\Pi)\equiv S\le(\Hcal(\lambda,a,b) ,{\p \Hcal(\lambda,a,b) \ov \del(\del_\rmx\phi)},{\p \Hcal(\lambda,a,b) \ov \del\Pi},\del_\rmx\phi,\Pi\ri)\,,
}
i.e. $\tilde S$ is the same function as $S$, but regarded as having the arguments $(a,b,\del_\rmx\phi,\Pi)$. As a first step towards obtaining $\del_\lambda \Hcal(\lambda,a,b)$, let us set $b=0$. Then
from what we said around \eqref{JxJbarx} it follows that
\es{SDef2}{
\del_\lambda \Hcal(\lam,a,0,\del_\rmx\phi,\Pi) = \tilde S(0,0,\del_\rmx\phi-2\pi a,\Pi+a/2)\,.
}
As discussed around \eqref{bguess}, obtaining such a closed form formula for $b\neq 0$ does not seem possible, but perturbation theory should be straightforward. Taking this as a hint, we expand the RHS of \eqref{SDef2} in $a$:
\es{StildeExp}{
\tilde S(0,0,\del_\rmx\phi-2\pi a,\Pi+a/2)=\sum_{m\geq 0} \frac{a^m}{m!}   D_1^m \tilde S(a,0,\del_\rmx\phi,\Pi) \,,
}
where the differential operator $D_1$ is defined by:
\es{D1def}{
  D_1 =-2\pi\biggl(\frac{\del }{\del(\del_\rmx\phi)} -{1\ov 4\pi} \frac{\del}{\del \Pi}\biggr) - \del_a \,. 
  }
The infinite series is easily seen to implement a translation $(-a,0,2\pi a, -a/2)$ on the arguments of $\tilde S$, proving \eqref{StildeExp}. Another nice way to see that the infinite series is equal to $\del_\lambda \Hcal(\lam,a,0)$, is to prove that they satisfy the same differential equation (regarded as an evolution equation with $a$ as time) with the same initial condition:
\es{DiffEqHS}{
D_1 \del_\lambda \Hcal(\lam,a,0,\del_\rmx\phi,\Pi) &= \del_\lambda \le(D_1 \Hcal(\lam,a,0,\del_\rmx\phi,\Pi) \ri)=0\,,\\
D_1 \sum_{m\geq 0} \frac{a^m}{m!}   D_1^m \tilde S(a,0,\del_\rmx\phi,\Pi) &= \le(-\sum_{m\geq 1} \frac{a^{m-1}}{(m-1)!} D_1^{m}+\sum_{m\geq 0}\frac{a^m}{m!}   D_1^{m+1}\ri) \tilde S(a,0,\del_\rmx\phi,\Pi)=0\,,\\
\del_\lambda \Hcal(\lam,0,0,\del_\rmx\phi,\Pi)&=\tilde S(0,0,\del_\rmx\phi,\Pi)\,,
}
where in the first line we used that $[D_1,\p_\lam]=0$ (because partial derivatives commute) and \eqref{JxJbarx} together with the expression of $J_x$ given in \eqref{ConsCurrComps}, while in the second line we relabeled the summation index to show that the two terms cancel. The third line is true by the definition of the $\lam$ deformation \eqref{SDef}.

The latter method generalizes to the $b\neq 0$ case. We want to find a differential operator $D_2=-\p_b+\dots$ that annihilates  $\p_\lam \Hcal$. 
The unique $D_2$ satisfying this requirements is: 
\es{D2def}{
  D_2 = -\le( \frac{\del \Hcal(\lam,a,b,\del_\rmx\phi,\Pi)}{\del(\del_\rmx\phi)} \frac{\del}{\del\Pi} +  \frac{\del \Hcal(\lam,a,b,\del_\rmx\phi,\Pi)}{\del\Pi} \frac{\del}{\del(\del_\rmx\phi)} \ri)- \del_b \,.
}
 To show that $D_2 \del_\lambda \Hcal=0$ we have to do some computations, as unlike in \eqref{DiffEqHS}, $[D_2,\p_\lam]\neq0$. We write
\es{D2dlamH}{
D_2 \del_\lambda \Hcal &=-\le(\p_b\p_\lam  \Hcal + \frac{\del \Hcal}{\del(\del_\rmx\phi)} \frac{\del \le(\del_\lambda \Hcal\ri) }{\del\Pi} +  \frac{\del \Hcal}{\del\Pi} \frac{\del \le(\del_\lambda \Hcal\ri) }{\del(\del_\rmx\phi)}\ri)\\
&=-\p_\lam\le( \p_b \Hcal + \frac{\del \Hcal}{\del(\del_\rmx\phi)} \frac{\del  \Hcal }{\del\Pi} \ri)=0\,,
}
where in the first line we wrote out the definitions, and in the second we commuted partial derivatives, and used that $\Hcal$ satisfies the differential equation as shown in \eqref{bguess}. Then we follow the same logic as in \eqref{DiffEqHS}. The proof of 
\es{D2Series}{
D_2 \sum_{n\geq 0} \frac{b^n}{n!}   D_2^n \tilde S(0,b,\del_\rmx\phi,\Pi) &=0
}
follows that in \eqref{DiffEqHS} verbatim. Thus the series and $\del_\lambda \Hcal$ satisfy the same evolution equation in $b$ (regarded as time), with the same initial conditions, given in the last line of \eqref{DiffEqHS}. It is easy to show that $[D_1,D_2]=0$, hence we can combine the two evolutions without encountering any issues, and we arrive at
\es{dellambdaHlambdaab}{
 \boxed{ \del_\lambda \Hcal(\lam,a,b,\del_\rmx\phi,\Pi)  = \sum_{m,n\geq 0} \frac{1}{m!n!} a^m b^n D_1^m D_2^n \tilde S(a,b,\del_\rmx\phi,\Pi) \,.}
}
 This is our key result, we will see that the infinite sum truncates in the cases of interest, and the resulting equation will allow us to write down an evolution equation for the energy levels.

To be able to deform by operators built from $\bar J$, we introduce a background field $\bar a$ that couples to it. As defined in \eqref{AnsatzEq}, the analogue of \eqref{JxJbarx} is $\frac{\del\Hcal}{\del \bar a} =  -i\bar J_\rmx$.  The corresponding unique differential operator that annihilates  $\del_\lambda \Hcal(\lam,a,\bar a , b)$ is given by: 
\es{D1bdef}{
  \bar D_1 = -2\pi\biggl(\frac{\del }{\del(\del_\rmx\phi)} +{1\ov 4\pi} \frac{\del}{\del \Pi}\biggr) - \del_{\bar a} \,. 
  }
   $\bar D_1$ commutes with $D_1,\, D_2$.
   
  Because we intend to build the deforming operator $S$ from $T_{\mu\nu}, \, J_\mu, \bar J_\mu$, it is a useful intermediate step to compute the action of the differential operators on these quantities. We collect the results in Table~\ref{tab:DAction}.
\begin{table}[!ht]
\begin{center}
  \begin{tabular}{@{} c||c|c|c|c|c|c|c|c @{}}
     & $J_t$ & $J_x$ & $\bar J_t$ & $\bar J_x$ &$T_{tt}$ & $T_{tx}$ & $T_{xt}$ & $T_{xx}$\\ 
    \hline\hline
    $D_1$& $2\pi $ & 0 & 0 & 0 & 0 & 0 & $iJ_t$ & $iJ_x$\\ 
\hline
    $\bar D_1$& 0 & 0 & $2\pi $  & 0 & 0 & 0 & $-i\bar J_t$ & $-i\bar J_x$ \\ 
\hline
     $D_2$&$i J_x$ & 0 & $i\bar J_x$ & 0 & $iT_{tx}$ & 0 & $-i(T_{tt}-T_{xx})$ & $-iT_{tx}$ 
  \end{tabular}
\caption{Action of the differential operators $D_1,\, \bar D_1,\,  D_2$ on the operators $ J_\mu,\, \bar J_\mu,\, T_{\mu\nu}$. \label{tab:DAction}}
\end{center}
\end{table}

As promised, we go through the $J\Tbar$ deformation in detail, and to shorten the discussion, we set $\bar a =0$. We will later write down the result including $\bar a \neq 0$  terms and also for the rest of the quadratic deformations. By the $J\Tbar$ deformation, we mean that we add to the Hamiltonian
\es{JTbarDef}{
S_{J\Tbar}&=2\pi  iJ_{[t|}T_{\zbar|x]}\\
&={\pi i\ov 2}\le(J_t T_{xx}+i J_t T_{tx}-J_x T_{xt}-iJ_x T_{tt}\ri)\,,
}
where the normalization is chosen so that we get $S_{J\Tbar}=J\Tbar$ in a CFT; our conventions are summarized in Appendix~\ref{app:conv}. We then compute all the non-vanishing derivatives (omitting $\bar D_1$):
\es{JTbarAllDeriv}{
D_1S_{J\Tbar}&=2\pi^2 i T_{\bar z x} \qquad D_2S_{J\Tbar}={\pi} J_{[t|} T_{t | x]}\\
D_1^2S_{J\Tbar}&=-\pi^2 J_x \qquad D_1 D_2S_{J\Tbar}=\pi^2 T_{t x}\,.
}
All other derivatives vanish. Using these formulas we conclude that
\es{JTbarFinal}{
  \del_\lambda \Hcal(\lambda,a,b) 
  = 2\pi  iJ_{[t|}T_{\zbar|x]} + {\pi} b J_{[t|} T_{t | x]}-{\pi^2 a^2 \ov 2} J_x+2\pi^2 i a T_{\bar z x} +  \pi^2 ab T_{t x} \,.
}
This universal equation holds for any Hamiltonian density in the class we considered. The term $ {\pi} b J_{[t|} T_{t | x]}$ is a linear combination of $J_\z\Tbar+J_\zbar\Thetabar$ and $J_\z\Theta+J_\zbar T$ deformations.
In a CFT the second deformation vanishes.

Now we are ready to systematize the derivation for all quadratic deformations that we can construct. The quadratic composite operators that obey factorization are
\es{QuadraticOps}{
``J\bar J"&\equiv - i J_{[t}\bar J_{x]}\\
``J\Tbar"&\equiv 2\pi  iJ_{[t|}T_{\zbar|x]}\\
``J \Theta"&\equiv -2\pi  i  J_{[t|}T_{z|x]}\\
``\bar J T"&\equiv -2\pi  i \bar J_{[t|}T_{z|x]}\\
``\bar J \bar\Theta"&\equiv 2\pi  i  \bar J_{[t|}T_{\bar z|x]}\\
``T\Tbar"&\equiv -2\pi^2 T_{t[t|}T_{x|x]}\,.
}
Instead of writing six long equations, we give ${\p\ov \p\lambda_\sO} \Hcal(\lambda,a,\bar a, b)$ with the deforming composite operator being $\sO$ in Table~\ref{tab:ddlam}. As promised, the equation is of the form \eqref{KeyEq}.
\begin{table}[!ht]
\begin{center}
\tabcolsep=0.08cm
\resizebox{1\textwidth}{!}{
  \begin{tabular}{@{} c||c|c|c|c|c|c|c|c|c|c|c|c|c|c @{}}
    \diagbox{$\sO$}{$+$} & $J\bar J$& $J\Tbar$ & $J \Theta$ & $\bar J T$ & $\bar J \bar\Theta$ & $T\Tbar$ & $J_t$ & $J_x$ & $\bar J_t$ & $\bar J_x$ &$T_{tt}$ & $T_{tx}$ & $T_{xt}$ & $T_{xx}$\\ 
    \hline\hline
    $J\bar J$ &  $1$ & $0$ & $0$ & $0$ & $0$ & $0$ & $0$ & $i \pi \bar a$ & $0$ & $-i\pi a$ & $0$ & $0$ & $0$ & $0$ \\ 
\hline
    $J\Tbar$ & $i \pi  \bar a$ & $1-\frac{b}{2}$ & $-\frac{b}{2}$ & $0$ & $0$ & $0$ & $0$ & $-\frac{\pi ^2}{2}  \left(a^2+\bar a^2\right)$ & $0$ & $\pi ^2 a \bar a$ & $0$ & $-\pi ^2 a (1-b)$ & $0$ & $i \pi ^2
   a$\\ 
\hline
  $J \Theta$    & $-i \pi  \bar a$ & $\frac{b}{2}$ & $1+\frac{b}{2}$ & $0$ & $0$ & $0$ & $0$ & $\frac{\pi ^2}{2}  \left(a^2+\bar a^2\right)$ & $0$ & $-\pi ^2 a \bar a$ & $0$ & $-\pi ^2 a (1+b)$ & $0$ & $-i
   \pi ^2 a$ \\
   \hline
   $\bar J T$     & $-i \pi  a$ & $0$ & $0$ & $1+\frac{b}{2}$ & $\frac{b}{2}$ & $0$ & $0$ & $\pi ^2 a \bar a$ & $0$ & $-\frac{\pi ^2}{2}  \left(a^2+\bar a^2\right)$ & $0$ & $-\pi ^2 \bar a (1+b)$ & $0$ & $-i \pi
   ^2 \bar a$\\
   \hline
     $\bar J \bar\Theta$  & $ i \pi  a$ & $0$ & $0$ & $-\frac{b}{2}$ & $1-\frac{b}{2}$ & $0$ & $0$ & $-\pi ^2 a \bar a$ & $0$ & $\frac{\pi ^2}{2}  \left(a^2+\bar a^2\right)$ & $0$ & $-\pi ^2 \bar a (1-b)$ & $0$ & $i \pi
   ^2 \bar a$\\
    \hline
     $T\Tbar$  & $0$ & $-i \pi  a$ & $-i \pi  a$ & $i \pi  \bar a$ & $i \pi  \bar a$ & $1$ & $0$ & $0$ & $0$ & $0$ & $0$ & $i \pi ^3 \left(a^2-\bar a^2\right)$ & $0$ & $0  $
  \end{tabular}}
\caption{The equation for ${\p\ov \p\lambda_\sO} \Hcal(\lambda,a,\bar a, b)$ can be read out from the this table as follows. The deforming operator $S=\sO$ labels the rows. We have to add up the operators in the top row with coefficients in the row labelled by $\sO$. For comparison, the $J\Tbar$ example for $\bar a=0$ is given in \eqref{JTbarFinal} in more conventional form.  \label{tab:ddlam}}
\end{center}
\end{table}

We will postpone the solution of these equations. Instead, we convert them into equations describing the evolution of the spectrum. In Appendix~\ref{app:recover} then we explain how to recover the classical Hamiltonian (and Lagrangian) from the solution of the spectrum.

\subsection{Flow equation for the spectrum}\label{sec:SpectrumEq}

The flow equations for the Hamiltonian density, \eqref{JTbarFinal} and Table~\ref{tab:ddlam}, can now be turned into a flow equation for the energy eigenvalues following the strategy outlined in Section~\ref{sec:strategy}: for a given eigenstate $\ket{n}$, we take the diagonal matrix element of the (conjectured) operator equation, for the composite operators use factorization, and replace the matrix elements that we encounter with:
\begin{equation}\label{AllMtxElements}
  \begin{aligned}
& & & \mathllap{\braket{n}{{\p\ov \p\lambda_\sO} \Hcal(\lambda,a,\bar a, b)}{n} }
= \mathrlap{ {\p\ov \p\lambda_\sO} E_n(\lambda,a,\bar a, b)\,, } \\
\braket{n}{J_{t}}{n}&={Q_n\ov L}\,, &  \braket{n}{J_x}{n}&=-{i\p_a E_n \ov L}\,, & \braket{n}{\bar J_{t}}{n}&={\bar Q_n\ov L}\,, &  \braket{n}{\bar J_x}{n}&={i\p_{\bar a} E_n \ov L}\\
\braket{n}{T_{tt}}{n}&=-{E_n\ov L}\,, & \braket{n}{T_{tx}}{n}&={i\p_b E_n \ov L}\,, & \braket{n}{T_{xt}}{n}&={iP_n\ov L}\,, &  \braket{n}{T_{xx}}{n}&=-{\p_L E_n}\,.
\end{aligned}
\end{equation}
For the time component of currents, $J_{t},\, \bar J_{t},\, T_{tt},\, T_{xt}$ the above equations follow from the definition of charge given in \eqref{ConsQuant} and \eqref{ConsCurr}. We coupled the spatial components of the currents to background fields, see \eqref{JxJbarx} and  \eqref{bguess}, thereby modifying the Hamiltonian, and we use the Hellmann-Feynman theorem $\braket{n}{{\de_\lam} H}{n}={\p\ov \p\lam}E_n$ to determine their matrix elements. The same logic is used to determine the first line of \eqref{AllMtxElements}.
The matrix element of $T_{xx}$ is curious, we obtain $-{\p_L E_n}$ from its interpretation as pressure. 
 From our perspective, the length of the spatial $S^1$ can be regarded as a {\it background field} on the same footing as $a,\bar a, b$, and from this point of view it becomes natural that its diagonal matrix element is obtained by taking a $\p_L$ derivative.  
 
 Executing this straightforward, but tedious task, we arrive at the differential equation describing the flow of energy eigenvalues. We again put the equations in a table, see Table~\ref{tab:energyflow}. For ease of reading, we write out the equation for the $J\Tbar$ deformation explicitly:
 \es{ExEq}{
 0=&{2L\ov i\pi}{\p\ov \p \lam_{J\Tbar}}E_n+\le(-\bar a  \hat{\bar Q }_n-\pi  (a^2-\bar a^2) L-(1-b) E_n+P_n\ri)\p_a E_n\\
   &-\bar a  \hat{Q }_n\p_{\bar a} E_n+(1-b) \hat{Q }_n\p_{b} E_n+L \hat{Q }_n\p_L E_n\,,\\
   \hat{Q}\equiv& Q+2\pi a L\,, \qquad \hat{\bar Q}\equiv \bar Q+2\pi \bar a L\,.
 }
 \begin{table}[!ht]
\begin{center}
\tabcolsep=0.08cm
  \begin{tabular}{@{} c||c|c|c|c|c @{}}
    \diagbox{$\sO$}{$+$} & ${\p\ov \p \lam_\sO}E_n$& $\p_a E_n$ & $\p_{\bar a} E_n$ & $\p_{b} E_n$ & $\p_L E_n$  \\ 
    \hline\hline
    $J\bar J$ & ${2L}$ & $-\hat{\bar Q }_n$ & $-\hat{Q }_n$ & $0$ & $0$ \\ 
\hline
    $J\Tbar$ & ${2L\ov i\pi}$ & \shortstack{$-\bar a  \hat{\bar Q }_n-\pi  (a^2-\bar a^2) L$\\$-(1-b) E_n+P_n$} & $-\bar a  \hat{Q }_n$ & $(1-b) \hat{Q }_n$ & $L \hat{Q }_n$\\ 
\hline
  $J \Theta$  & ${2L\ov i\pi}$ & \shortstack{$\bar a  \hat{\bar Q }_n+\pi  (a^2-\bar a^2) L$\\$-(1+b) E_n-P_n$} & $\bar a  \hat{Q }_n$ & $(1+b) \hat{Q }_n$ & $-L \hat{Q }_n$\\
   \hline
   $\bar J T$  & $-{2L\ov i\pi}$ & $-a \hat{\bar Q }_n$ & \shortstack{$-a \hat{Q }_n+\pi  (a^2-\bar a^2) L$\\$-(1+b) E_n-P_n$} & $-(1+b) \hat{\bar Q }_n$ & $L \hat{\bar Q }_n$\\
   \hline
     $\bar J \bar\Theta$ & $-{2L\ov i\pi}$ & $a \hat{\bar Q }_n$ & \shortstack{$a \hat{Q }_n-\pi  (a^2-\bar a^2) L$\\$-(1-b) E_n+P_n$} & $-(1-b) \hat{\bar Q }_n$ & $-L \hat{\bar Q }_n$ \\
    \hline
     $T\Tbar$  & $-{L\ov \pi^2}$ & $a E_n$ & $\bar a  E_n$ & \shortstack{$-a \hat{Q }_n+\bar a  \hat{\bar Q }_n$\\$\pi  (a^2-\bar a^2) L-P_n$} & $-E_n L$
  \end{tabular}
\caption{The flow equation for the energy eigenvalue can be read out from the this table as follows. The deforming operator $S=\sO$ labels the rows. We have to add up the terms in the top row with coefficients in the row labelled by $\sO$ and equate it to zero. For reference, the second line is given in conventional form in \eqref{ExEq}, where we also define $\hat{Q}, \hat{\bar Q}$. \label{tab:energyflow}}
\end{center}
\end{table}
 
The main power of our method comes from its ability to solve theories, where we consider a linear combination of irrelevant deformations. Recall, that as reviewed in Section~\ref{sec:strategy} the case of $T\Tbar$ deformation of a relativistic QFT can be solved without introducing the background fields $a,\bar a, b$ \cite{Smirnov:2016lqw,Cavaglia:2016oda}, while the $J\Tbar$ (or equivalently the $\bar J T$) deformation can be solved using holomorphy \cite{Chakraborty:2018vja}. However, the combination of $T\Tbar$ and  $J\Tbar$ leads to the loss of both Lorentz invariance and holomorphy, and the aforementioned methods do not apply. Hence let us introduce a length scale $\ell$ with $[\ell]=-1$ and real dimensionless couplings $g_\sO$:
\es{Dimless}{
\lam_{J\Tbar}&\equiv i g_{J\Tbar} \ell\,, \qquad \lam_{J \Theta}\equiv i g_{J \Theta} \ell\,,\qquad \lam_{\bar J T}\equiv -i g_{\bar J T} \ell\,, \qquad \lam_{\bar J \bar\Theta}\equiv -i g_{\bar J \bar\Theta} \ell\,,\\
\lam_{T\Tbar}&\equiv  g_{T\Tbar} \ell^2\,.
}
By changing $\ell$, we obtain a one-parameter family of theories. Note that because $J\bar J$ is a marginal operator it is not included among the deforming operators. The energy levels evolve according to the equation:
\es{FullFlow}{
L{\p\ov \p \ell}E_n={\pi g_{J\Tbar}\ov 2} \text{II} + {\pi g_{J\Theta}\ov 2} \text{III}+{\pi g_{\bar J T}\ov 2} \text{IV}+{\pi g_{\bar J  \bar\Theta}\ov 2} \text{V}+{2\pi^2 \ell \,g_{T\Tbar}} \text{VI}\,,
}
where the Roman numerals stand for one row of Table~\ref{tab:energyflow} (omitting the ${\p\ov \p \lam_\sO}E_n$ entry). We note that a similar equation can also be obtained at the level of the operator equations included in Table~\ref{tab:ddlam}. We will solve \eqref{FullFlow} in Section~\ref{sec:Solution} with the initial conditions determined in Section~\ref{sec:lindef}. 

\subsection{Fixing ambiguities in the initial conditions}\label{sec:AmbiguityFix}

In Section~\ref{sec:Ambiguity}, we discussed some ambiguities in the initial conditions. These ambiguities are fixed by the form of the conserved currents that we gave in \eqref{ConsCurrComps}. Conversely, only using conservation, in \eqref{ConsCurrComps} the $x$ components of currents could be shifted in the same  way as in \eqref{ShiftsByIdentity}. Since we have the additional scale $\ell$ with $[\ell]=-1$ in the problem, the ambiguities could be made even more severe. We have to invoke additional principles to fix them.

Let us start with $T_{xx}$, from which we want to require $ \braket{n}{T_{xx}}{n}=-{\p_L E_n}$, see \eqref{AllMtxElements}. The Noether stress tensor given  in \eqref{ConsCurrComps} achieves this. Since we have not written down $T^\text{(min)}_{xx}$ there, we omit the details and just state that there indeed exists a shift involving the background fields that makes  the expression of $T_{xx}$ in \eqref{AllMtxElements} match with $T^\text{(min)}_{xx}$. 

Coupling a scalar theory to a constant background gauge field $A_\mu=(a,0)$ in the  Hamiltonian formalism amounts to the shift $\Hcal(a)=h(\del_\rmx\phi-2\pi a,\Pi+a/2)$. This was already used above, see \eqref{JxJbarx}. Gauge invariance forbids the addition of $A_\mu A^\mu$ terms. At $\lam_{\sO}=\ell=0$, we determined the Hamiltonian in Appendix~\ref{app:scalar}, in \eqref{DeformedHam}. Comparing this to the algebraic result $T^\text{(min)}_{tt}$, we require the shift:
\es{TttShifts}{
T_{tt}&=T^\text{(min)}_{tt}-{\pi a^2\ov 1-b}-{\pi \bar a^2\ov 1+b}\,,
}
and the shifts of of $J^\text{(min)}_x,\, \bar J^\text{(min)}_x,\, T^\text{(min)}_{tx}$ follow from these shifts according to \eqref{ShiftsByIdentity}. We have checked that these shifts are exactly the ones needed to reproduce the currents given in \eqref{ConsCurrComps}. Integrating \eqref{TttShifts} according to the rule \eqref{ConsQuant}, we get
\es{HShift}{
H(a,\bar a, b)=\frac{H^{(0)}+bP^{(0)}}{1-b^2}+ {a Q^{(0)}+\pi a^2 L\ov 1-b}+ {\bar a \bar Q^{(0)}+\pi \bar a^2 L\ov 1+b}\,,
}
where we used \eqref{AllCharges}.

After settling the ambiguities, we are ready to give the initial conditions for the energy flow equations. Because the operators in \eqref{HShift} commute, we can easily convert it to an expression for the energy eigenvalues:
\es{EnergyInit}{
E_n=\braket{n}{H(a,\bar a, b)}{n}=\frac{E_n^{(0)}+bP_n}{1-b^2}+ {a Q_n+\pi a^2 L \ov 1-b}+ {\bar a \bar Q_n+\pi \bar a^2 L\ov 1+b}\,.
}
We will use this as initial data for the flow equation \eqref{FullFlow} in Section~\ref{sec:Solution}.

We remark that the algebraic approach does not break down without the additional requirements discussed in this section. E.g. we could define $J_x=2\pi i\le(\frac{\del \sH}{\del(\del_\rmx\phi)} -{1\ov 4\pi} \frac{\del \sH}{\del \Pi}\ri)+2ia g_1(b)$, which would in turn lead to the modification of entries in Tables~\ref{tab:DAction},~\ref{tab:ddlam},~\ref{tab:energyflow}, and ultimately lead to a different (and uglier) \eqref{FullFlow}. The solution would also change, but setting the background fields to zero must give an identical result for the energy spectrum of the theory deformed by quadratic composite operators. 

\section{Checks}\label{sec:checks}

\subsection{A classical field theory check}\label{sec:CLcheck}

In the previous section we conjectured a set of universal equations, \eqref{JTbarFinal} and Table~\ref{tab:ddlam}, governing the evolution of the Hamiltonian under irrelevant deformations based on the classical free scalar with shift symmetry. In this section, we check a restriction of these equations to the case of one conserved $U(1)$ current which can generate a symmetry other than shifts, and a much more general classical scalar theory.

Consider a collection of scalars $\phi_I$ and momenta $\Pi^I$ and Hamiltonian density $H=h(\phi_I,\del_x\phi_I,\Pi^I)$, for example  scalars with a potential or a sigma model.
We sum over repeated $I,J,\dots$ indices.
The Hamilton equations of motion are:
\es{HamEq}{
  \del_t\phi_I = -i \frac{\del \sH}{\del\Pi^I}\,, \qquad
  \del_t\Pi^I = i \le(\frac{\del \sH}{\del\phi_I} -  \del_x \frac{\del \sH}{\del(\del_x\phi_I)} \ri)\,.
}
For our sign conventions refer to \eqref{HamEoM}. The theory is translation invariant and in complete analogy to \eqref{ConsCurrComps} the components of the conserved stress tensor are:
\es{ConservedTmunu}{
    T_{tt} & = - \sH \,, \qquad T_{tx}  = - i \frac{\del \sH}{\del\Pi^I} \frac{\del \sH}{\del(\del_x\phi_I)} \,, \\
    T_{xt} & = - i \Pi^I \del_x \phi_I \,,\qquad T_{xx}  =   \frac{\del \sH}{\del(\del_x\phi_I)} \del_x\phi_I + \frac{\del \sH}{\del\Pi^I} \Pi^I- \sH \,.
}
If the Hamiltonian is invariant under some continuous symmetry group acting like $\delta\phi_I=\Lambda_I(\phi)$ and $\delta\Pi^I=-\Pi^J\frac{\del\Lambda_J}{\del\phi_I}$, it has a conserved current
\es{ConsK}{
  K_t = \Pi^I \Lambda_I \,, \qquad
  K_x = i \frac{\del 
  \sH}{\del(\del_x\phi_I)} \Lambda_I \,.
}
For the familiar case of the $O(2)$ symmetric scalar field, we have $\Lambda_I =\ep_{IJ} \phi_J$.  For the shift symmetry we discussed in Section~\ref{sec:Classical}, $\Lambda =4\pi$ and the current $K$ of \eqref{ConsK} corresponds to  the difference of holomorphic and antiholomorphic currents $K_\mu=J_\mu-\bar J_\mu$, hence we chose a different name for it. 

We want to understand linear deformations by coupling to the background fields $a$ and $b$ according to the rules ${\p\sH\ov\p a}=iK_x,\, {\p\sH\ov\p b}=-i T_{tx}$. Following the strategy of Section~\ref{sec:Classical} to write down a flow equation for $\sH(\lam,a,b)$, we want to find commuting differential operators $\sD_1,\, \sD_2$ that act on functions of variables $(a,b,\phi_I,\del_x\phi_I,\Pi^I)$ and  that annihilate $\p_\lam \sH$. This is possible, and their expressions are:
\es{sDops}{
\sD_1&=-\Lam_I {\p\ov \del(\del_x\phi_I)}-\p_a\,,\\
\sD_2&=- \le( \frac{\del \sH}{\del(\del_x\phi_I)} \frac{\del }{\del\Pi^I}+\frac{\del \sH}{\del\Pi^I} \frac{\del }{\del(\del_x\phi_I)}\ri)-\p_b\,.
}
 It is now straightforward to compute the results in Table~\ref{tab:DAction2}. Note that in the case of the scalar with shift symmetry investigated in Section~\ref{sec:Classical}, $\bar a=a$ and $\sD_1=D_1+\bar D_1$. The results in this table are in complete agreement with those in Table~\ref{tab:DAction}, if we remember that $K_\mu=J_\mu-\bar J_\mu$ and $\sD_1=D_1+\bar D_1$.
\begin{table}[!ht]
\begin{center}
  \begin{tabular}{@{} c||c|c|c|c|c|c @{}}
     & $K_t$ & $K_x$ & $T_{tt}$ & $T_{tx}$ & $T_{xt}$ & $T_{xx}$\\ 
    \hline\hline
    $\sD_1$& $0 $ & 0 & 0 & 0 & $iK_t$ & $iK_x$\\ 
\hline
     $\sD_2$&$i K_x$ & 0 &  $iT_{tx}$ & 0 & $-i(T_{tt}-T_{xx})$ & $-iT_{tx}$ 
  \end{tabular}
\caption{Action of the differential operators $\sD_1,\,  \sD_2$ on the operators $ K_\mu, \, T_{\mu\nu}$. \label{tab:DAction2}}
\end{center}
\end{table}

Since everything in Section~\ref{sec:Classical} followed from the results of  Table~\ref{tab:DAction}, and we recovered those results in this more general setting, we reach the same conclusions as in the rest of that section. We conclude that we found additional evidence for the universality of the equations collected in Table~\ref{tab:ddlam}.

 To be explicit we summarize how to read off the results appropriate for the case at hand. Besides $T\bar T$, we can only consider the deformation by 
\es{Kdefs}{
\sO_1&\equiv2\pi  iK_{[t|}T_{\zbar|x]} \stackrel{\text{(free scalar)}}{=}J\Tbar-\bar J\bar\Theta\,,\\
\sO_2&\equiv2\pi  iK_{[t|}T_{z|x]} \stackrel{\text{(free scalar)}}{=} \bar J T- J\Theta\,.
}
Then using also that $\bar a=a$, Table~\ref{tab:ddlam} collapses to Table~\ref{tab:ddlam2}. We obtained the latter table both from Table~\ref{tab:ddlam} using the rules explained and also by direct computation. Notably, only the quadratic composite operators make an appearance, and the linear operators are absent. 
\begin{table}[!ht]
\begin{center}
\tabcolsep=0.08cm
  \begin{tabular}{@{} c||c|c|c @{}}
    \diagbox{$\sO$}{$+$} & $\sO_1$& $\sO_2$ &  $T\Tbar$ \\ 
    \hline\hline
    $\sO_1$ &  $1-\frac{b}{2}$ & $-\frac{b}{2}$ & $0$ \\ 
\hline
   $\sO_2$ & $\frac{b}{2}$ & $1+\frac{b}{2}$ & $0$ \\
   \hline
        $T\Tbar$  & $-i \pi  a$ & $-i \pi  a$ & $1$ 
  \end{tabular}
\caption{The equation for ${\p\ov \p\lambda_\sO} \Hcal(\lambda,a, b)$ can be read off from the table in exactly the same way as from Table~\ref{tab:ddlam}.  \label{tab:ddlam2}}
\end{center}
\end{table}

Continuing in this direction, we could obtain a flow equation for the spectrum in the same way as in Section~\ref{sec:Classical}.  We do not write down the result of this straightforward exercise here.  As, unlike in the case of the deformed free scalar with shift symmetry, we do not have a point in the parameter space with a CFT with (anti)holomorphic currents, which was crucial in determining the initial conditions in Section~\ref{sec:lindef}, we do not know how to determine the initial conditions for neither flow equations.
This is the reason, we only presented the treatment of the more general case as a check on the conjectured universality of the flow equations. The initial conditions could however be obtained in Gaussian theories: the massive complex boson and fermion, and it is an interesting future direction to obtain the spectrum of their irrelevant deformations.

\subsection{A perturbative quantum check}\label{sec:PertCheck}

The universal equations~\eqref{JTbarFinal} and Table~\ref{tab:ddlam} for the Hamiltonian density $\Hcal=-T_{tt}$ can be checked in quantum perturbation theory around a CFT, order by order in~$\lambda$ and exactly in the linear couplings $a$, $\bar{a}$ and~$b$.
These equations are statements about local operators modulo derivative terms, because they involve collision limits that are only defined up to derivatives.

In line with the rest of the paper we place the theory on $S^1\times\R$ and work in the Hamiltonian formalism and on a fixed time slice.\footnote{Translation to the path integral formalism should be straightforward.}
We expand all local operators in Fourier modes.
For example, the CFT's holomorphic stress-tensor is
\begin{equation}\label{Virasoroshifted}
  T(x) = - \biggl(\frac{2\pi}{L}\biggr)^2 \sum_{k=-\infty}^\infty e^{2\pi ikx/L} \, \ell_k , \qquad
  [\ell_k , \ell_m] = (k-m)\ell_{k+m} + \frac{c}{12} k^3 \delta_{k+m,0} ,
\end{equation}
in terms of shifted Virasoro modes $\ell_k\equiv L_k-\delta_{k,0}\,c/24$.
See Appendix~\ref{app:qupertformulas} for more conventions and explicit formulas.
All operators of interest are constructed from the dimensionless modes $\ell_k$, $\bar{\ell}_k$, $j_k$, $\bar{j}_k$ of the CFT stress-tensor $T(x)$, $\Tbar(x)$ and two independently-conserved currents $J(x)$, $\Jbar(x)$.

Schematically, one proceeds as follows.
First turn on~$\lambda$.
In our formalism, $T_{xt}$, $J_t$, $\Jbar_t$ are fixed.  Once the mode expansions of $J_\mu$, $\Jbar_\mu$ and $T_{\mu\nu}$ are known up to order $\lambda^{p-1}$, one computes the quadratic operator by which to deform, for example the collision limit $``J\Tbar"=2\pi i J_{[t|}T_{\bar{z}|x]}$, to deduce $T_{tt}$ hence $H=-\int dx\,T_{tt}$ to order $\lambda^p$.  Then conservation gives $\del_x T_{tx}$, $\del_x T_{xx}$, $\del_x J_x$, $\del_x\Jbar_x$ thus gives all modes of $T_{tx}$, $T_{xx}$, $J_x$, $\Jbar_x$ except their zero modes (since $\del_x e^{inx}$ vanishes for $n=0$).  Locality fixes these zero modes up to ambiguities explained in Section~\ref{sec:Ambiguity}: shifts by multiples of the identity.
Then $a$, $\bar{a}$, $b$ are turned on using the same steps.

The rest of the section spells out details.
We introduce useful deformations of the modes $\ell_k$, $j_k$, $\bar{\ell}_k$, $\bar{j}_k$ in Section~\ref{sssec:spectrumgenerating}.
Next, we tackle the two key difficulties: finding OPEs such as $2\pi i J_{[t|}T_{\bar{z}|x]}$ in Section~\ref{sssec:computingOPEs},
and finding zero modes of $T_{tx}$, $T_{xx}$, $J_x$, $\Jbar_x$ in Section~\ref{sssec:zeromodes}.
Section~\ref{sssec:summary} summarizes all the steps needed to do perturbation theory in our setting.

For the $J\Tbar$ deformation we performed calculations specified by the procedure up to order~$\lambda^2$, with $\bar{a}=0$ and exactly in $a$, $b$, and confirmed the universal equation.  At this order quantum effects could have spoiled the equation but some coefficients cancel.  Let us see why quantum effects arise at this order.  Our quantum calculations reduce to classical calculations by replacing all commutators by Poisson brackets, replacing all collision limits of operators by (coincident-point) products of functions, and setting $c=0$.  The last requirement comes from comparing the commutators $[T(x),T(y)]$ and
\begin{equation}\label{TbarTbarcommutatorCFT}
  [\Tbar(x),\Tbar(y)] = - 2\pi i \biggl(\frac{c}{12} \delta'''(x-y) + 2 \Tbar(y) \delta'(x-y) - \del_y \Tbar(y) \delta(x-y)\biggr)
\end{equation}
to their classical Poisson bracket analogues which have no $(c/12)\delta'''(x-y)$ term.
Quantum perturbation theory expresses $T_{\mu\nu}(x)$ and $J_\mu(x)$ as series in~$\lambda$ of sums of composite operators built from the CFT operators $T(x)$, $J(x)$, $\Tbar(x)$.
Dimensional analysis restricts the set of operators that can appear.
We are interested in terms multiplying~$c$.
Factors of~$c$ appear in commutators~\eqref{TbarTbarcommutatorCFT} multiplied by the distribution $\delta'''(x-y)$, which involves two additional derivatives and one fewer stress-tensor compared to other terms.
In expressions of $J_\mu$ and~$T_{\mu\nu}$, operators multiplying~$c$ thus involve two derivatives.
For $T_{tt}$, dimensional analysis only allows $\del_x^2 J$ at order~$\lambda$, and $\del_x^2T$, $\del_x^2\Tbar$, $J\del_x^2 J$, $\del_x J\del_x J$ at order~$\lambda^2$ (the term $\del_x^2T$ is actually not produced by commutators).
Since the universal equation is defined modulo derivatives, derivative terms $\del_x^2 J$ and $\del_x^2\Tbar$ cannot spoil it.
However, the terms $J\del_x^2 J$ and $\del_x J\del_x J$ could arise with different ($b$-dependent) coefficients, thus fail to give a derivative.  These terms would then affect energy levels.
The outcome of our calculation is that the terms have equal coefficients so that they combine as $J\del_x^2 J+\del_x J\del_x J = \del_x^2 J^2/2$.
The universal equation is thus confirmed, as are the energy levels.

Note that this check is not subsumed in the comparison of $J\Tbar$-deformed energy levels with earlier literature.  Indeed, these previous results were worked out by imposing holomorphy of~$J_\mu$ (our definitions of $J_\mu$ differ slightly, as discussed in Appendix~\ref{app:JTbar}) which cannot be done once we turn on backgrounds $a$ and~$b$.

\subsubsection{Spectrum-generating operators}\label{sssec:spectrumgenerating}

We now return to a general quadratic and linear deformations constructed from $T_{\mu\nu}$, $J_\mu$, $\Jbar_\mu$, and we introduce operators $\Lambda_k$, $\Upsilon_k$, $\overline{\Lambda}_k$, $\overline{\Upsilon}_k$ that play an important role when computing OPEs later.\footnote{In the case of the $J\Tbar$ deformation, the operators $\overline{\Lambda}_k$ should reduce to effectively non-local state-dependent Virasoro generators found previously in~\cite{Guica:2017lia,Bzowski:2018pcy,Chakraborty:2018vja,Nakayama:2018ujt}.}
For brevity we choose notations adapted to deformations by a single quadratic operator, with a single coupling~$\lambda$, but it is easy to generalize.
We call ``eigenstate'' or ``state in the spectrum'' a joint eigenstate of the various conserved charges: energy~$H$, momentum~$P$ and charges $Q$, $\overline{Q}$.

Tracking the $\lambda$-dependence of eigenstates is impractical because one must determine how each eigenstate $\ell_{k_1}\dots j_{m_1}\dots\bar{\ell}_{n_1}\dots\bar{j}_{p_1}\dots\ket{\text{primary}}$ in the CFT evolves.
Instead we track relations between these states.  More precisely we construct perturbatively a family of operators (see~\eqref{explicitLambdaetc} for $O(\lambda)$ terms)
\begin{equation}
  \Lambda_k=\ell_k+O(\lambda) , \qquad \Upsilon_k=j_k+O(\lambda) , \qquad \overline{\Lambda}_k=\bar{\ell}_k+O(\lambda) , \qquad \overline{\Upsilon}_k=\bar{j}_k+O(\lambda)
\end{equation}
that generate the spectrum in the sense that acting on an eigenstate gives another eigenstate.
These operators can be defined abstractly as the result of ``conjugating'' the original modes $\ell_k$, $j_k$, $\bar{\ell}_k$, $\bar{j}_k$ by the deformation.
For any eigenstate $\ket{n}_\lambda$ that is the image of some CFT state~$\ket{n}$ under the deformation,
$\Lambda_k\ket{n}_\lambda$ is defined as the image of $\ell_k\ket{n}$ under the deformation, and likewise $\Upsilon_k\ket{n}_\lambda \equiv (j_k\ket{n})_\lambda$ and $\overline{\Lambda}_k\ket{n}_\lambda \equiv (\bar{\ell}_k\ket{n})_\lambda$ and $\overline{\Upsilon}_k\ket{n}_\lambda \equiv (\bar{j}_k\ket{n})_\lambda$ are images of $j_k\ket{n}$, $\bar{\ell}_k\ket{n}$, $\bar{j}_k\ket{n}$ under the deformation.\footnote{More precisely, the CFT spectrum has states with degenerate energy and momentum and charge, for instance $\ell_{-4}\ket{0}$ and $\ell_{-2}^2\ket{0}$, and to disentangle $(\ell_{-4}\ket{0})_\lambda$ from $(\ell_{-2}^2\ket{0})_\lambda$ one uses KdV conserved charges, under which the CFT spectrum is non-degenerate.
  For the $J\Tbar$ flow, and for any of Zamolodchikov's antisymmetric quadratic combinations of conserved currents,
these further conserved charges also exist in the deformed theory.}

This abstract definition does not help compute $\Lambda_k$, $\Upsilon_k$, $\overline{\Lambda}_k$, $\overline{\Upsilon}_k$ but leads to various properties.
\begin{itemize}
\item Given a state $\ket{n}=\ket{h,q,\bar{h},\bar{q}}$ in the CFT with $\ell_0$, $j_0$, $\bar{\ell}_0$, $\bar{j}_0$ eigenvalues $h$, $q$, $\bar{h}$, $\bar{q}$ respectively, its image under the flow obeys $\Lambda_0\ket{n}_\lambda = (\ell_0\ket{n})_\lambda = h\ket{n}_\lambda$ and so on.  In that sense, $\Lambda_0\pm\overline{\Lambda}_0$, $\Upsilon_0$, $\overline{\Upsilon}_0$ acting on $\ket{n}_\lambda$ measure the energy, momentum and charges of the original state~$\ket{n}$.
\item Since charge and momentum of states are fixed, $\Upsilon_0=j_0$, $\overline{\Upsilon}_0=\bar{j}_0$,  and $\Lambda_0-\overline{\Lambda}_0=\ell_0-\bar{\ell}_0$.
\item The operators obey the same Virasoro and Ka\v{c}--Moody algebra as $\ell_k$, $j_k$, $\bar{\ell}_k$, $\bar{j}_k$, namely $[\Lambda_k,\overline{\Lambda}_m] = [\Lambda_k,\overline{\Upsilon}_m] = [\Upsilon_k,\overline{\Lambda}_m] = [\Upsilon_k,\overline{\Upsilon}_m] = 0$ and
  \begin{equation}\label{sameVirasoro}
    \begin{aligned}
      [\Lambda_k , \Lambda_m] & = (k-m)\Lambda_{k+m} + \frac{c}{12} k^3 \delta_{k+m,0} ,
      & [\Lambda_k , \Upsilon_m] & = -m \Upsilon_{k+m} ,
      & [\Upsilon_k, \Upsilon_m] & = k \delta_{k+m} , \\
      [ \overline{\Lambda}_k , \overline{\Lambda}_m ] & = (k-m)\overline{\Lambda}_{k+m} + \frac{c}{12} k^3 \delta_{k+m,0} ,
      & [\overline{\Lambda}_k , \overline{\Upsilon}_m] & = -m \overline{\Upsilon}_{k+m} ,
      & [\overline{\Upsilon}_k, \overline{\Upsilon}_m] & = k \delta_{k+m} .
    \end{aligned}
  \end{equation}
\item Acting with $\Lambda_k$ or $\Upsilon_k$ or $\overline{\Lambda}_k$ or $\overline{\Upsilon}_k$ on an eigenstate~$\ket{n}_\lambda$ gives another eigenstate.  Its energy is higher than that of $\ket{n}_\lambda$ if $k<0$ and lower if $k>0$.  One could call these operators ``raising'' or ``lowering'' operators according to the sign of~$k$, but importantly their existence does not make the spectrum trivial.  Indeed, energies of different eigenstates are shifted by different amounts.
\end{itemize}
Explicit low-order perturbative calculations suggest a last property for our class of deformations.
\begin{itemize}
\item The Hamiltonian $H$ can be written as a function $\mathsf{H}(\lambda;\Lambda_0,\Upsilon_0,\overline{\Lambda}_0,\overline{\Upsilon}_0)=\frac{2\pi}{L}(\Lambda_0+\overline{\Lambda}_0)+O(\lambda)$, given explicitly for the $J\Tbar$ deformation in~\eqref{JTbarHamiltonianAsLambda}.
  In particular, eigenstates of $H$, $P$, $Q$, $\overline{Q}$ are the same as eigenstates of
  $\Lambda_0$, $\Upsilon_0$, $\overline{\Lambda}_0$, $\overline{\Upsilon}_0$.
  From it we deduce that the energy of a state $\ket{h,q,\bar{h},\bar{q}}_\lambda$ is $\mathsf{H}(\lambda;h,q,\bar{h},\bar{q})$ since
\begin{equation}
  H \ket{h,q,\bar{h},\bar{q}}_\lambda = \mathsf{H}(\lambda;\Lambda_0,\Upsilon_0,\overline{\Lambda}_0,\overline{\Upsilon}_0) \ket{h,q,\bar{h},\bar{q}}_\lambda
  = \mathsf{H}(\lambda;h,q,\bar{h},\bar{q}) \ket{h,q,\bar{h},\bar{q}}_\lambda .
\end{equation}
Energy levels then depend on the original energy, momentum and charges in the same was as the Hamiltonian depends on $\Lambda_0\pm\overline{\Lambda}_0$, $\Upsilon_0$, $\overline{\Upsilon}_0$.
Reversing the logic, our solution~\eqref{BrutalFormula} for energy levels thus predicts the exact Hamiltonian.
For example for the $J\Tbar$-deformed CFT with $a=\bar{a}=b=0$, we get
\begin{equation}\label{HpredictJTbar}
  H \overset{\text{prediction}}{=} \frac{2 \pi}{L} \biggl( \Lambda_0 -\overline{\Lambda}_0
  - \frac{L^2}{2\pi^4\lambda^2} \biggl( 1-\frac{2\pi^2i\lambda}{L} \Upsilon_0-\sqrt{\bigl(1- 2\pi^2i(\lambda/L) \Upsilon_0\bigr)^2 - 2 \bigl(2\pi^2i\lambda/L\bigr)^2 \overline{\Lambda}_0} \biggr) \biggr) .
\end{equation}
\end{itemize}

How do we find the expressions of the spectrum-generating operators $\Lambda_k$, $\Upsilon_k$, $\overline{\Lambda}_k$, $\overline{\Upsilon}_k$ order-by-order in~$\lambda$ in terms of the CFT modes $\ell_k$, $j_k$, $\bar{\ell}_k$, $\bar{j}_k$?
The construction of $\del_\lambda\Lambda_k$, $\del_\lambda\Upsilon_k$, $\del_\lambda\overline{\Lambda}_k$, $\del_\lambda\overline{\Upsilon}_k$ is easiest to do in terms of the spectrum-generating operators themselves; it can then be translated to the CFT modes using expressions of $\Lambda_k$, $\Upsilon_k$, $\overline{\Lambda}_k$, $\overline{\Upsilon}_k$ at the previous order in~$\lambda$.
While we eventually do all of our calculations in terms of $\Lambda_k$, $\Upsilon_k$, $\overline{\Lambda}_k$, $\overline{\Upsilon}_k$,
derivatives with respect to couplings always denote derivatives at fixed $\ell_k$, $j_k$, $\bar{\ell}_k$, $\bar{j}_k$.
This makes it a bit awkward to reconstruct an operator $\Ocal=\sum_{n\geq 0} \frac{1}{n!} \lambda^n \Ocal^{(n)}$ from its $\lambda$~derivative because the $\lambda^p/p!$ term in $\del_\lambda\Ocal$ works out to be
\begin{equation}\label{Ocalnplus1}
  (\del_\lambda\Ocal)^{(p)} = \Ocal^{(p+1)} + \sum_{n=0}^p \binom{p}{n} \bigl( \del_\lambda\bigl(\Ocal^{(n)}\bigr) \bigr)^{(p-n)}
\end{equation}
where we expanded $\del_\lambda\bigl(\Ocal^{(n)}\bigr)$ in powers of~$\lambda$.
Indeed, derivatives of $\Lambda_k$, $\Upsilon_k$, $\overline{\Lambda}_k$, $\overline{\Upsilon}_k$ that appear in $\del_\lambda\bigl(\Ocal^{(n)}\bigr)$ are themselves series in~$\lambda$ when expressed in terms of these spectrum-generating operators.

In practice we first note that $\del_\lambda\Upsilon_0=\del_\lambda\overline{\Upsilon}_0=\del_\lambda(\Lambda_0-\overline{\Lambda}_0)=0$ by charge and momentum conservation.  Then we construct $\del_\lambda\overline{\Lambda}_0=\del_\lambda\Lambda_0$ such that~\eqref{HpredictJTbar} holds (or its analogue for other deformations).
We find it by solving~\eqref{HpredictJTbar} for $\overline{\Lambda}_0$ in terms of $H$ and $\Upsilon_0$ and $\Lambda_0-\overline{\Lambda}_0$,
\begin{equation}\label{Lambdabar0}
  \overline{\Lambda}_0
  =
  \frac{1}{2}
  \biggl(1- \frac{2\pi^2i\lambda}{L} \Upsilon_0\biggr)\biggl(  \frac{HL}{2\pi} - \Lambda_0 + \overline{\Lambda}_0\biggr)
  + \frac{\pi^4\lambda^2}{2L^2} \biggl(\frac{HL}{2\pi} - \Lambda_0 + \overline{\Lambda}_0\biggr)^2 ,
\end{equation}
and taking a $\del_\lambda$ derivative.
It is straightforward to show that $\del_\lambda \overline{\Lambda}_0$ commutes with $\Lambda_0-\overline{\Lambda}_0$, namely all terms $\Lambda_{m_1}\dots\Upsilon_{n_1}\dots\overline{\Lambda}_{\bar{m}_1}\dots\overline{\Upsilon}_{\bar{n}_1}\dots$ in $\del_\lambda\overline{\Lambda}_0$ obey $\sum m+\sum n=\sum\bar{m}+\sum\bar{n}$.
At the orders we checked we additionally find that there are no terms that commute with $\Lambda_0$ (or equivalently with $\overline{\Lambda}_0$), namely no term with $\sum m+\sum n=\sum\bar{m}+\sum\bar{n}=0$.

This observation is essential for the following construction to work.

We now know $\del_\lambda\Lambda_0=\del_\lambda\overline{\Lambda}_0$ up to a certain order in~$\lambda$ and want to construct other $\del_\lambda\Lambda_k$, $\del_\lambda\Upsilon_k$, $\del_\lambda\overline{\Lambda}_k$, $\del_\lambda\overline{\Upsilon}_k$ that are consistent with the commutators~\eqref{sameVirasoro}.
For instance we want to preserve $[\overline{\Lambda}_0,\Lambda_k]=[\overline{\Lambda}_0,\Upsilon_k]=0$.  From their derivatives we learn that we need
\begin{equation}\label{Lambdabar0dlambdaUpsilonk}
  [\overline{\Lambda}_0,\del_\lambda\Lambda_k]=[\Lambda_k,\del_\lambda\overline{\Lambda}_0] , \qquad \text{and} \qquad
  [\overline{\Lambda}_0,\del_\lambda\Upsilon_k]=[\Upsilon_k,\del_\lambda\overline{\Lambda}_0] .
\end{equation}
Crucially, the right-hand sides do not contain any term of the form $\Lambda\dots\Upsilon\dots\overline{\Lambda}_{m_1}\dots\overline{\Upsilon}_{n_1}\dots$ with $\sum\bar{m}+\sum\bar{n}=0$, because as we mentioned, $\del_\lambda\overline{\Lambda}_0$ do not contain such terms.
Then \eqref{Lambdabar0dlambdaUpsilonk} fixes $\del_\lambda\Lambda_k$ and $\del_\lambda\Upsilon_k$ up to such terms, and we choose to define $\del_\lambda\Lambda_k$ and $\del_\lambda\Upsilon_k$ without any such term, even though we could add terms $\Lambda\dots\Upsilon\dots\overline{\Lambda}_{m_1}\dots\overline{\Upsilon}_{n_1}\dots$ with $\sum\bar{m}+\sum\bar{n}=0$ without spoiling~\eqref{Lambdabar0dlambdaUpsilonk}.

We define $\del_\lambda\overline{\Lambda}_k$ and $\del_\lambda\overline{\Upsilon}_k$ similarly, based on $[\overline{\Lambda}_0,\overline{\Lambda}_k] = -k\overline{\Lambda}_k$, which gives $[\overline{\Lambda}_0,\del_\lambda\overline{\Lambda}_k]+k\del_\lambda\overline{\Lambda}_k = [\overline{\Lambda}_k,\del_\lambda\overline{\Lambda}_0]$
hence fixes $\del_\lambda\overline{\Lambda}_k$ up to terms of the form $\Lambda_{m_1}\dots\Upsilon_{n_1}\dots\overline{\Lambda}_{\bar{m}_1}\dots\overline{\Upsilon}_{\bar{n}_1}\dots$ with $\sum\bar{m}+\sum\bar{n}=k$.
We choose to define $\del_\lambda\overline{\Lambda}_k$ without any such term.
Equivalently, these terms are characterized by $\sum m+\sum n=0$, so this is really the analogue of the condition we put on terms appearing in $\del_\lambda\Lambda_k$ and $\del_\lambda\Upsilon_k$.

These definitions reduce for $k=0$ to the ones we already imposed.

Finally we must check our constructed $\del_\lambda\Lambda_k$, $\del_\lambda\Upsilon_k$, $\del_\lambda\overline{\Lambda}_k$, $\del_\lambda\overline{\Upsilon}_k$ give rise to the commutators~\eqref{sameVirasoro}.  Let us just show one calculation explicitly: that $\del_\lambda\bigl([\Lambda_k,\Lambda_m]-(k-m)\Lambda_{k+m}-k^3\delta_{k+m,0}\,c/12\bigr)$ vanishes.
First, note that this derivative is built from some $\del_\lambda\Lambda_n$, which by construction have no terms that commute with~$\overline{\Lambda}_0$, so it is enough to check that $[\overline{\Lambda}_0,\del_\lambda(\dots)]$ vanishes.
We compute (at an order in~$\lambda$ where we know the commutators~\eqref{sameVirasoro} but not yet their $\lambda$ derivative)
\begin{equation}
  \begin{aligned}
    [\overline{\Lambda}_0,\del_\lambda(\dots)]
    & = [\overline{\Lambda}_0, [\del_\lambda \Lambda_k, \Lambda_m]]
    + [\overline{\Lambda}_0, [\Lambda_k, \del_\lambda \Lambda_m]]
    -(k-m)[\overline{\Lambda}_0, \del_\lambda\Lambda_{k+m}]
    \\
    & = [[\overline{\Lambda}_0, \del_\lambda \Lambda_k], \Lambda_m]
    + [\Lambda_k, [\overline{\Lambda}_0, \del_\lambda \Lambda_m]]
    -(k-m)[\overline{\Lambda}_0, \del_\lambda\Lambda_{k+m}]
    \\
    & = [[\Lambda_k, \del_\lambda \overline{\Lambda}_0], \Lambda_m]
    + [\Lambda_k, [\Lambda_m, \del_\lambda \overline{\Lambda}_0]]
    -(k-m)[\Lambda_{k+m}, \del_\lambda\overline{\Lambda}_0]
    \\
    & = \bigl[[\Lambda_k,\Lambda_m] - (k-m) \Lambda_{k+m}, \del_\lambda \overline{\Lambda}_0\bigr] = 0 .
  \end{aligned}
\end{equation}

This concludes the construction of spectrum-generating operators $\Lambda_k$, $\Upsilon_k$, $\overline{\Lambda}_k$, $\overline{\Upsilon}_k$.
At each order in~$\lambda$ one should check that $\del_\lambda\overline{\Lambda}_0$ deduced from~\eqref{Lambdabar0} has no term $\Lambda\dots\Upsilon\dots\overline{\Lambda}\dots\overline{\Upsilon}\dots$ that commutes with $\Lambda_0$.
Other properties of these operators then come for free.

\subsubsection{Computing OPEs}\label{sssec:computingOPEs}
The Virasoro (and Ka\v{c}--Moody) algebras \eqref{Virasoroshifted} and~\eqref{appEVirasoroKacMoody} obeyed by $\ell_k$, $j_k$, $\bar{\ell}_k$, $\bar{j}_k$ are unchanged by the deformation, and the same is true for commutators of local operators such as $T(x)$ whose expression in terms of modes does not depend on couplings.

On the other hand, OPEs of such coupling-independent operators change.\footnote{This is a somewhat different situation than the OPEs considered in~\cite{Aharony:2018vux}, because what these authors denote $T,\Theta,\Thetabar,\Tbar$ are certain components of the deformed stress-tensor $T_{\mu\nu}$, whereas here we consider OPEs, in the deformed theory, of the CFT operators.}
For instance,
\begin{equation}\label{JTbarOPEdeformed}
  J(x) \Tbar(y)
  = 2\pi \lambda \biggl( \frac{c/2}{(x-y)^4}
  + \frac{2\Tbar(y)}{(x-y)^2}
  + \frac{\del_y \Tbar(y)}{(x-y)} \biggr)
  + O((x-y)^0) + O(\lambda^2)
\end{equation}
in the $J\Tbar$-deformed theory, even though the left-hand side has no $\lambda$~dependence whatsoever.
In our formalism, this seemingly contradictory result comes from how the notion of well-defined operator depends on~$\lambda$.
In the CFT,
\begin{equation}\label{JTbarmodesinCFT}
  J(x) \Tbar(x) = i \biggl( \frac{2\pi}{L} \biggr)^3 \sum_{k,m} e^{2\pi i(k+m)x/L} j_k \bar{\ell}_m
\end{equation}
is well-defined, in the sense that each mode ($k+m=\text{constant}$) is an infinite sum that truncates when acting on any state in the spectrum.\footnote{By ``state in the spectrum'' we mean an eigenstate of the Hamiltonian, momentum, and conserved charge.}  The spectrum depends on~$\lambda$ and in the deformed theory the sum fails to truncate, so that the coincident-point operator $J(x)\Tbar(x)$ is ill-defined.
The correct OPE~\eqref{JTbarOPEdeformed} can be checked in principle by comparing matrix elements ${}_\lambda\braket{n}{J(x)\Tbar(y)}{n'}_\lambda$ between eigenstates $\ket{n}_\lambda$, $\ket{n'}_\lambda$ to those of the right-hand side.

Let us briefly discuss collision limits in a CFT when working explicitly in modes.
First consider $J(x)J(y)$.  To get a finite collision limit we reorder modes using the commutator (we set $L=2\pi$ to shorten expressions)
\begin{equation}\label{JJsuperExplicitCFTOPE}
  \begin{aligned}
    J(x)J(y) & = - \sum_{k,m} e^{i(kx+my)} j_k j_m
    = - \sum_{k>m} e^{i(kx+my)} [j_k, j_m] - \sum_{k,m} e^{i(kx+my)} \nop{j_k j_m} \\
    & = \frac{1}{(2\sin\frac{x-y}{2})^2} - \sum_{k,m} e^{i(kx+my)} \nop{j_k j_m}
    = \frac{1}{(x-y)^2} + \frac{1}{12} + \nop{J(y)J(y)} + O(x-y)
  \end{aligned}
\end{equation}
where $\nop{j_k j_m} = (j_k j_m \text{ if } k<m \text{ else } j_m j_k)$.\footnote{The shift by $1/12$ is the expected shift $\ell_n=L_n-(c/24)\delta_{n,0}$ once one remembers that $\nop{J^2}$ is twice the Sugawara stress-tensor, which has central charge $c=1$ in this case.}
The reason $- \sum_{k,m} e^{i(k+m)x} \nop{j_k j_m}$ has finite matrix elements in any energy eigenstate $\ket{n}$ of the CFT is that $\nop{j_kj_m}\ket{n}$ vanishes for large enough $k$ or~$m$ (thanks to the normal-ordering) while $\bra{n'}\nop{j_kj_m}$ vanishes for negative enough $k$ or~$m$.  Altogether only finitely many $k$ and~$m$ can contribute to a given $\braket{n'}{\nop{j_kj_m}}{n}$.
For more complicated examples such as collisions of Sugawara stress-tensors $\frac{1}{2}\nop{J^2}$,
the prescription is still to reorder modes using commutators until modes are all ordered, then evaluate the series such as $\sum_k e^{ik(x-y)} k$ that arise.
Normal-ordered products have finite collision limits.
In a CFT, the shortcut to get the regularized collision limits of operators $\nop{J^p}$ is simply to normal-order the modes and take $x=y$.

Consider now the collision limit of a product $\mathcal{A}(x)\mathcal{B}(y)$ of local operators in the deformed theory.\footnote{The deformed operators $J_\mu(x)$, $\Jbar_\mu(x)$, $T_{\mu\nu}(x)$ are eventually built from various collision limits at~$x$ of the CFT operators $J$, $\Jbar$, $T$, $\Tbar$ and their derivatives, so commutators of two such operators at different points $x_1$ and~$x_2$ vanish, namely these operators are still local after deformations.  The fact that the $T\Tbar$ deformation preserves locality was already observed in~\cite{Aharony:2018vux}.}

We can apply a similar idea: express $\mathcal{A}$ and $\mathcal{B}$ in terms of spectrum-generating operators $\Lambda_k$, $\Upsilon_k$, $\overline{\Lambda}_k$, $\overline{\Upsilon}_k$ then sort these operators by increasing~$k$.
Let us call the resulting normal-ordered product $\operatorname{Sort}(\mathcal{A}(x)\mathcal{B}(y))$.
Since this places lowering operators to the right of raising ones, all matrix elements in energy eigenstates truncate the sums to finitely many terms, hence remain finite as $x\to y$.  The $x\to y$ collision limit $\operatorname{Sort}(\mathcal{A}\mathcal{B})$ is thus well-defined.
Unfortunately, this ordering prescription is not consistent with locality, namely we find by explicit calculations that the commutator of $\operatorname{Sort}(\mathcal{A}\mathcal{B})(y)$ with a local operator at $w$ fails to vanish for $w\neq y$.

To preserve locality we cannot use the shortcut of normal ordering.
Instead, we keep track of all commutators when reordering the operators $\Lambda_k$, $\Upsilon_k$, $\overline{\Lambda}_k$, $\overline{\Upsilon}_k$ as we did in~\eqref{JJsuperExplicitCFTOPE} in the CFT case.
Once all terms are ordered, the coefficient of each product $\Upsilon\dots\Lambda\dots\overline{\Lambda}\dots\overline{\Upsilon}\dots$, often an infinite sum, should be evaluated and expanded as $x\to y$.
The sought-after collision limit is then the $(x-y)^0$ term.
Besides the normal-ordered product $\operatorname{Sort}(\mathcal{A}\mathcal{B})$ it may include additional terms, for example the shift by $1/12$ in~\eqref{JJsuperExplicitCFTOPE}.

We  obtained the non-trivial OPE~\eqref{JTbarOPEdeformed} of the CFT local operators $J(x)$ and~$\Tbar(y)$ by following these steps in the $J\Tbar$-deformed theory.  Converting from modes $j_k$ and $\bar{\ell}_k$ to operators $\Lambda_k$, $\Upsilon_k$, $\overline{\Lambda}_k$ uses (the inverse of) the explicit formulas~\eqref{explicitLambdaetc}.
At order~$\lambda$, $J(x)\Tbar(y)$ includes terms such as $\sum_{k,m,n} ({\dots}) \Upsilon_k \Upsilon_m \overline{\Lambda}_n$ in which the $\Upsilon$ must be reordered.  The commutator terms $[\Upsilon_k,\Upsilon_m]\overline{\Lambda}_n$ give sums of modes $\overline{\Lambda}_n$ whose coefficients are singular as $x\to y$, which lead to $\Tbar$ and $\del_y\Tbar$ terms in~\eqref{JTbarOPEdeformed}.  The $c$-dependence in the OPE comes directly from the $c$-dependence of the dictionary~\eqref{explicitLambdaetc} between CFT modes and deformed ones $\Lambda_k$, $\Upsilon_k$, $\overline{\Lambda}_k$.

The OPE of $J(x)\Tbar(y)$ is only one term in the OPE $2\pi i J_{[t|}T_{\bar{z}|x]}$ that we are really interested in, because the components $J_\mu$ and $T_{\mu\nu}$ depend on~$\lambda$.
Among other terms, $J_x$ contains $\lambda\Tbar$, whose OPE with $\Tbar$ cancels most terms in~\eqref{explicitLambdaetc}.  Altogether, the collision limit we care about works out to be
\begin{equation}\label{JtTbarxOPE}
  2\pi i J_{[t|}(x) T_{\bar{z}|x]}(y) = - \lambda \frac{\pi \del_y \Tbar(y)}{x-y} + O((x-y)^0) + O(\lambda^2) .
\end{equation}
The operator $\del_y\Tbar(y)$ is a derivative, as expected from general considerations about antisymmetric combinations of conserved currents.
We are actually interested in the $(x-y)^0$ term in this OPE\@.  Working it out we get a finite collision limit expressed in terms of $\Lambda_k$, $\Upsilon_k$, $\overline{\Lambda}_k$.
By definition, $\del_\lambda T_{tt} = -2\pi i J_{[t|}T_{\bar{z}|x]}$, where the $\lambda$ derivative is taken at fixed $\ell_k$, $j_k$, $\bar{\ell}_k$, so to get the next power of $\lambda$ in~$T_{tt}$ one must either translate $2\pi i J_{[t|}T_{\bar{z}|x]}$ to the modes $\ell_k$, $j_k$, $\bar{\ell}_k$, or account for non-zero $\del_\lambda\Lambda_k$, $\del_\lambda\Upsilon_k$, $\del_\lambda\overline{\Lambda}_k$.

\subsubsection{Locality by turning on linear deformations}\label{sssec:zeromodes}

Once $T_{tt}$~is known, conservation equations give $\del_x J_x$, $\del_x\Jbar_x$, $\del_x T_{tx}$, $\del_x T_{xx}$, but give no information on zero modes of these spatial components of currents.
Finding the zero modes is absolutely crucial because they affect all modes of quadratic products such as $2\pi i J_{[t|}T_{\bar{z}|x]}$, used to define $T_{tt}$ at the next order in~$\lambda$.
In principle one should impose locality to find these modes, namely one should ask for the commutator with CFT operators $T$, $J$, $\Tbar$, $\Jbar$ to be zero at separated points.  This is very difficult: if we work in terms of $\Lambda_k$, $\Upsilon_k$, $\overline{\Lambda}_k$, $\overline{\Upsilon}_k$ then commutators with modes of $T$, $J$, $\Tbar$, $\Jbar$ are complicated; and if we work in terms of $\ell_k$, $j_k$, $\bar{\ell}_k, \bar{j}_k$ there is no good way to determine whether a given sum of products of modes is well-defined, as we discussed near~\eqref{JTbarmodesinCFT}.

To get around this hurdle, and to turn on $a$ and~$b$, we treat our classical evolution equation~\eqref{JTbarFinal} or Table~\ref{tab:ddlam} as providing an Ansatz for $T_{tt}(\lambda,a,\bar{a},b)$ hence for $J_x=i\del_a T_{tt}$ and $\Jbar_x=-i\del_{\bar{a}} T_{tt}$ and $T_{tx}=-i\del_bT_{tt}$ and $T_{xx}=\del_L(L T_{tt})$.
As everything else in this subsection, checking the Ansatz is done order by order in~$\lambda$, so let us assume that $T_{tt}(\lambda,a,\bar{a},b)$ is known up to order $\lambda^{p-1}$, and exactly in $a$,~$\bar{a}$,~$b$.

The order $\lambda^p$ term of $T_{tt}$ provided by our classical equation is correct for $a=\bar{a}=b=0$ by definition of the deformation.  Then, to show that $T_{tt}(\lambda,a,\bar{a},b)$ matches the definition of the $a$, $\bar{a}$, $b$ deformations, we need only check that for any $(a,\bar{a},b)$ the derivatives $i\del_a T_{tt}$, $-i\del_{\bar{a}} T_{tt}$ and $-i\del_b T_{tt}$ are indeed equal to the correct components $J_x$, $\Jbar_x$, $T_{tx}$ defined by the conservation equations and by locality.
One has to check at each order in~$\lambda$ that the Ansatz obeys conservation, using explicit expressions for a given deformation, as we do not have a general proof of conservation.
On the other hand, locality is automatic because $T_{tt}$ is constructed from local operators (including collisions, computed as explained above), and taking $a$, $\bar{a}$, $b$ derivatives commutes with taking a commutator with the reference local (CFT) operators $T$, $J$, $\Tbar$, $\Jbar$.
Local conserved currents $(J_t,J_x)$ and $(J_t,i\del_a T_{tt})$ with the same time components can only differ by multiples of the identity, so our Ansatz gives the correct $a$, $\bar{a}$, $b$ deformations up to this ambiguity, already discussed in Section~\ref{sec:Ambiguity}.

\subsubsection{Summary of the procedure}\label{sssec:summary}

To start the whole process we need to know the ``initial data'': $J_\mu$, $\Jbar_\mu$, $T_{\mu\nu}$ at order $\lambda^0$ for all $a$, $\bar{a}$, $b$ (and~$L$).  At this order, the stress-tensor and conserved current are linear combinations~\eqref{CurrentGuess} of the CFT ones.
The dependence on $a$, $\bar{a}$, $b$ is fixed up to the ambiguities~\eqref{ShiftsByIdentity} under shifts by multiples of the identity.
We also keep $\Lambda_k=\ell_k+O(\lambda)$ etc.\ with no $a$ nor~$b$ dependence at order~$\lambda^0$.

One safe way to avoid accidentally writing ill-defined products such as~\eqref{JTbarmodesinCFT} is to work in terms of the spectrum-generating operators $\Lambda_k$, $\Upsilon_k$, $\overline{\Lambda}_k$, $\overline{\Upsilon}_k$ and systematically commute operators with larger~$k$ towards the right of any product.
We use \eqref{Ocalnplus1} in the form $\Ocal^{(p+1)} = (\del_\lambda\Ocal)^{(p)} - \cdots$ to deduce an operator at order $\lambda^{p+1}$ from its derivative at order~$\lambda^p$.

The concrete procedure to get the order~$\lambda^{p+1}$ terms in $J_\mu$, $\Jbar_\mu$, $T_{\mu\nu}$ (hence in~$H$) knowing all order~$\lambda^p$ terms is then as follows.
\begin{enumerate}
\item \label{step:collision} Determine up to order~$\lambda^p$ the collision limit $2\pi i J_{[t|}T_{\bar{z}|x]}+\pi bJ_{[t|}T_{t|x]}+\dots$ appearing on the right-hand side of~\eqref{JTbarFinal} or its generalizations from Table~\ref{tab:ddlam}.  For $a=\bar{a}=b=0$ this is $\del_\lambda T_{tt}$, while for nonzero $(a,\bar{a},b)$ it is only an Ansatz, checked later.
  To deal with derivative ambiguities, one includes with unknown coefficients the derivative of every local operator allowed by dimensional analysis.
\item \label{step:dlambdaLambda} Write the expression for $\del_\lambda\overline{\Lambda}_0$ given in~\eqref{Lambdabar0} or its generalizations including linear deformations.\break
  \textbf{Check} that this Ansatz is valid, in that it produces no terms that commute with~$\Lambda_0$ (or equivalently with~$\overline{\Lambda}_0$).  Deduce all other $\del_\lambda\Lambda_k$, $\del_\lambda\Upsilon_k$, $\del_\lambda\overline{\Lambda}_k$, $\del_\lambda\overline{\Upsilon}_k$ up to order~$\lambda^p$.
\item Use \eqref{Ocalnplus1} to deduce $T_{tt}^{(p+1)}$ from $(\del_\lambda T_{tt})^{(p)}$ computed in step~\ref{step:collision} and from the order $\lambda^{p-k}$ terms of derivatives $\del_\lambda(T_{tt}^{(k)})$, $0\leq k\leq p$, which involve terms computed in step~\ref{step:dlambdaLambda}.  Deduce $H^{(p+1)}$.
  By construction of $\del_\lambda\overline{\Lambda}_0$ the Hamiltonian is expressed in terms of $\Lambda_0$, $\Upsilon_0$, $\overline{\Lambda}_0$, $\overline{\Upsilon}_0$.
\item Likewise, work out $\del_a$, $\del_{\bar{a}}$, $\del_b$ derivatives of $\Lambda_k$, $\Upsilon_k$, $\overline{\Lambda}_k$, $\overline{\Upsilon}_k$ up to order $\lambda^{p+1}$ by noticing that their $\del_\lambda$ derivative is known to order~$\lambda^p$.  For instance $\del_\lambda\del_a\Lambda_k = \del_a\del_\lambda\Lambda_k = \del_a(\text{known}+O(\lambda^{p+1}))$, which only requires knowing $\del_a$ derivatives of $\Lambda_k$, $\Upsilon_k$, $\overline{\Lambda}_k$, $\overline{\Upsilon}_k$ up to order~$\lambda^p$.
\item Compute $J_x=i\del_aT_{tt}$ and $\Jbar_x=-i\del_{\bar{a}}T_{tt}$ and $T_{tx}=-i\del_bT_{tt}$ and $T_{xx}=\del_L(L T_{tt})$ up to~$\lambda^p$.\break
  \textbf{Check} current conservation $[H,J_t]+\del_x J_x=0$ and similarly for $\Jbar_\nu$ and $T_{\mu\nu}$.
  This check fixes the unknown derivative terms in step~\ref{step:collision}.
  As explained in Section~\ref{sssec:zeromodes}, locality is automatic and the check proves that the Ansatz for~$T_{tt}^{(p+1)}$ is correct.
\end{enumerate}

We performed this procedure (without $\Jbar$) to obtain order $\lambda^2$ terms in $J_\mu$ and $T_{\mu\nu}$ for the $J\Tbar$ deformation, thus checking the universal equation~\eqref{JTbarFinal} as a local operator equation modulo derivatives.  This proves to order $\lambda^2$ that the spectrum, including linear backgrounds, is exactly as predicted by the evolution equation~\eqref{ExEq}, which we solve exactly in~\eqref{JTbarSol} and~\eqref{BrutalFormula}.

The procedure is quite bulky, and needs to be simplified before it can be pushed to much higher order.
Perhaps the path integral formalism can help, but one would have to carefully work out OPEs such as~\eqref{JTbarOPEdeformed} and~\eqref{JtTbarxOPE} in this approach.
One potential difficulty is that the spectrum-generating operators $\Lambda_k$, $\Upsilon_k$, $\overline{\Lambda}_k$, $\overline{\Upsilon}_k$ seem to be more natural in the Hamiltonian formalism than in the path integral.

In this discussion we worked with spectrum-generating operators $\Lambda_k$, $\Upsilon_k$, $\overline{\Lambda}_k$, $\overline{\Upsilon}_k$ to make writing normal ordered products easier, but we nevertheless considered $\ell_k$, $j_k$, $\bar{\ell}_k$, $\bar{j}_k$ as fixed when defining derivatives such as~$\del_\lambda$.
Just as one switches from the Schr\"odinger to the Heisenberg picture in quantum mechanics,
we could switch from having $\lambda$-dependent states $\ket{n}_\lambda$, hence $\lambda$-dependent spectrum-generating operators, to having a $\lambda$-independent spectrum and spectrum-generating operators.
This would mean $\del_\lambda\Lambda_k$ and so on would vanish, while $\del_\lambda\ell_k$ etc.\ would not vanish any longer.
It may be instructive to translate our universal equation to this picture.

\section{The spectrum from the solution of flow equations}\label{sec:Spectrum}

\subsection{Solving a large family of theories}\label{sec:Solution}

\eqref{FullFlow} is a rather intimidating nonlinear PDE of five variables, nevertheless we will write down a closed form solution of it. While we suspect that there should be a straightforward derivation of the solution, we obtained the solution below using intuition from known results and solving the equations in series form. Before presenting the solution, we sketch the steps that led us to it. 

In our conventions (with $g_{J\Tbar}=1$ or  $\lam_{J\Tbar}=i \ell$, see \eqref{Dimless}), the spectrum of the $J\Tbar$-deformed CFT in \cite{Chakraborty:2018vja}:
\es{JTbarSol}{
E_n L&={1\ov \pi^3\le(\ell\ov L\ri)^2}\le(1+\pi Q \le(\ell\ov L\ri)+\pi^3 p\,  \le(\ell\ov L\ri)^2- \sqrt{1+2\pi Q \le(\ell\ov L\ri)-\le(2\pi^3(\ep_0-p)-\pi^2 Q^2  \ri)\le(\ell\ov L\ri)^2} \ri)\,,\\
p&\equiv P_nL\,, \qquad \ep_0\equiv E_n^{(0)} L\,.
}
The spectrum of the $T\Tbar$-deformed CFT is (with $g_{T\Tbar}=1$ or $\lam_{T\Tbar}=\ell^2$) \cite{Smirnov:2016lqw,Cavaglia:2016oda}:
\es{TTbarSol}{
E_n L&={1\ov 2\pi^2 \le(\ell\ov L\ri)^2}\le(1- \sqrt{1-4\pi^2 \ep_0 \,\le(\ell\ov L\ri)^2+4\pi^4p^2\,\le(\ell\ov L\ri)^4} \ri)\,.
}
The spectrum can be checked to obey the Burgers equation \eqref{Burgers}, if we use the relation $\lam_\text{\eqref{Burgers}}=-2\pi^2 \ell^2$ as explained around  \eqref{factor}. Based on these examples, a reasonable guess is that the spectrum of the full theory with all couplings turned on may take the form:
\es{SpectrumGuess}{
E_n L&={1\ov \# \le(\ell\ov L\ri)^2}\le(P_2\le({\ell\ov L}\ri)- \sqrt{P_4\le({\ell\ov L}\ri)} \ri)\,,
}
where $P_n\le({\ell\ov L}\ri)$ denotes the an $n$th order polynomial of ${\ell\ov L}$. The coefficients of the polynomials can depend on the initial data $\ep_0,p,Q,\bar Q$ and the (generalized) background fields $a,\bar a, b,L$. We require that as ${\ell\ov L}\to 0$ we recover the initial condition given in \eqref{EnergyInit}. By matching to a high order series solution of \eqref{FullFlow}, the Ansatz \eqref{SpectrumGuess} can be verified and the coefficients determined.

The string construction to be discussed in Section~\ref{sec:String} suggests additional structure, namely that $E_n$ is a solution of a quadratic equation. Following this hint, the most compact form that we managed to bring the solution to is:
\es{BrutalFormula}{
0&=A (\ep_n-s)^2-B(\ep_n-s)+C\,,\\
\ep_n &\equiv E_n \hat{L} =s+{B-\sqrt{B^2-4AC}\ov 2A}\,,\\
\hat{L}&\equiv (1-b^2)L\,, \quad \mu\equiv {\pi \ell\ov \hat{L}}\,, \quad  \hat{Q}\equiv Q+2\pi a L\,, \quad \hat{\bar Q}\equiv \bar Q+2\pi \bar a L\,, \quad \sA\equiv aL\, \quad \bar \sA\equiv \bar a L\,,\\
G_{J\Tbar}&\equiv (1-b)g_{J\Tbar}\,, \quad G_{J \Theta}\equiv (1+b)g_{J \Theta}\,, \quad G_{\bar J T}\equiv (1+b)g_{\bar J T}\,,\quad G_{\bar J \bar\Theta}\equiv (1-b)g_{\bar J \bar\Theta}\,,\\
\hat G_{T\Tbar}&\equiv G_{T\Tbar}+\frac{1}{2} \pi  (G_{J\Tbar} G_{J \Theta}+G_{\bar J T} G_{\bar J \bar\Theta})\,, \quad G_{T\Tbar}\equiv (1-b^2)g_{T\Tbar}\,,\\
A&=\Bigl(\frac{\pi}{2}  \bigl(G_{J\Tbar}^2 +G_{\bar J T}^2\bigr)+\hat G_{T\Tbar}\Bigr)\mu ^2 \\
B&=1  +\bigl(G_{J\Tbar} \hat{Q}+G_{\bar J T}
   \hat{\bar Q}\bigr) \mu+ B_2 \mu ^2\\
 C&= - \Bigl(G_{J\Tbar} \hat{Q} (\ep_0-p)+G_{\bar J T} \hat{\bar Q} (\ep_0+p)
   +2 \sA G_{\bar J T} \hat{\bar Q} \bigl(\hat{Q}-\pi\sA\bigr)
   +2 \bar\sA G_{J\Tbar} \bigl(\hat{\bar Q}-\pi\bar\sA\bigr) \hat{Q}
   \Bigr)\mu+C_2\mu ^2\\
 s&= \ep_0+bp +(1+b)\sA  \hat{Q} +(1-b)\bar\sA \hat{\bar Q}- \pi (1+b) \sA^2-\pi (1-b) \bar\sA^2\,.
}
where we defined the coefficients $B_2, C_2$ to make the expressions more digestible. They are given by:
\es{B2C2}{
  B_2={}& - \pi (\ep_0-p) G_{J\Tbar}^2 - \pi (\ep_0+p) G_{\bar JT}^2
  - 2\ep_0 \hat G_{T\Tbar}
  \\
  &-2 \sA(\hat{Q}-\pi\sA) \left(\pi  G_{\bar J T}^2+\hat G_{T\Tbar}\right)-2 \bar\sA(\hat{\bar Q}-\pi\bar\sA) \left(\pi  G_{J\Tbar}^2+\hat G_{T\Tbar}\right)\\
   C_2={}&
   \frac{\pi}{2} (\ep_0 - p)^2 G_{J\Tbar}^2
   + \frac{\pi}{2} (\ep_0 + p)^2 G_{\bar J T}^2
   + (\ep_0^2 - p^2) \hat G_{T\Tbar}
   \\
   & + 2\sA(\hat{Q}-\pi\sA) \bigl( \pi (\ep_0 + p)  G_{\bar J T}^2 + (\ep_0 - p) \hat G_{T\Tbar} \bigr)
   \\
   & + 2\bar\sA(\hat{\bar Q}-\pi\bar\sA) \bigl( \pi (\ep_0-p) G_{J\Tbar}^2 + (\ep_0 + p) \hat G_{T\Tbar} \bigr)\\
   &
   + 2\pi G_{\bar J T}^2 \sA^2 (\hat{Q}-\pi \sA)^2
   + 4 \sA (\hat{\bar Q}-\pi\bar\sA) \bar\sA (\hat{Q}-\pi\sA)\hat G_{T\Tbar}
   + 2\pi G_{J\Tbar}^2 \bar\sA^2 (\hat{\bar Q}-\pi\bar\sA)^2\,.
}
Let us highlight some properties of this  lengthy set of expressions. When $\mu=0$, corresponding to linear deformations of a CFT, $A=C=0$ and $B=1$, hence $\ep_n=s$. $s$ matches \eqref{EnergyInit}, once we account for the fact that $E_n=\ep_n/\hat L$, and this was the initial condition that we used to find the solution given in \eqref{BrutalFormula}. It is not surprising that by forming dimensionless combinations $\mu, \sA,\bar \sA$ we simplified formulas. Curiously, we managed to absorb all the $b$-dependence of $A,B,C$ into the redefined couplings $G_\sO$ and $\mu$, but even using these definitions $s$ still depends on $b$ explicitly. We also managed to absorb all the $G_{J \Theta},\, G_{\bar J \bar\Theta}$ dependence into a shifted $T\Tbar$ coupling,  $\hat G_{T\Tbar}$ defined in \eqref{BrutalFormula}. That $A,B,C,s$ are polynomials in $\ep_0,p,\sA,\bar \sA, \hat Q,\hat{\bar Q}$ is a consequence of \eqref{dellambdaHlambdaab} truncating at finite order.

We used the background fields mainly as an auxiliary device. If we are only interested in quadratic deformations, we can turn the background fields off: $\sA=\bar \sA=b=0,\, \hat Q=Q,\,\hat{\bar Q}=\bar Q,\, G_\sO=g_\sO$. In the absence of background fields, the expressions simplify significantly, it is only the first line of $B_2$ and $C_2$ that remain. The special cases of $J\Tbar$ and $T\Tbar$ deformations given in \eqref{JTbarSol} and \eqref{TTbarSol} are reproduced as special cases.

For large initial energy $\ep_0$ with the other quantum numbers fixed, the spectrum formally behaves as 
\es{Asymp}{
\ep_n&=\sqrt{-{\ep_0\ov \pi A}}+O(1)\,,\\
A&=\left(\frac{1}{2} \pi  (G_{J\Tbar}^2 +G_{\bar J T}^2)+G_{T\Tbar}+\frac{1}{2} \pi  (G_{J\Tbar} G_{J \Theta}+G_{\bar J T} G_{\bar J \bar\Theta})\right)\mu ^2\,,
}
where we repeated $A$ from \eqref{BrutalFormula} for convenience. For $A>0$ the energies become complex for some value of $\mu$. In \cite{McGough:2016lol} it was proposed for the $T\Tbar$ deformation that such states should be discarded from the spectrum, turning the theory into a quantum mechanical theory with finite number of states. The validity of this proposal is clearly beyond what can be assessed by the local field theory tools used in this paper. For $A<0$ the energies are real, and combined with the Cardy growth of the density of states this leads to Hagedorn growth of the density of states \cite{Dubovsky:2012wk,Giveon:2017nie}. For the pure $J\Tbar$ and $\bar J T$ deformations $A>0$, and the spectrum necessarily becomes complex \cite{Chakraborty:2018vja}. Once we turn on $T\Tbar$ (or equivalently $J \Theta$ in the presence of $J\Tbar$) with a sufficiently negative coupling constant, we can make $A<0$ and the asymptotic spectrum real. It could happen even for the $A<0$ case that for some intermediate energies the spectrum becomes complex.

Finally, it is natural to ask about the meaning of the other branch of the square root in  \eqref{BrutalFormula}, $\ep_n^{(+)}= s+{B+\sqrt{B^2-4AC}\ov 2A}$. For $\mu\to 0$ the energy of these states would diverge. Nevertheless, these ``eigenvalues'' play an interesting role in the modular differential equation that the torus partition function obeys in $T\Tbar$ and $J\Tbar$ deformed theories \cite{Aharony:2018bad,Aharony:2018ics}.

\subsection{Exploring the coupling space}\label{sec:Couplingspace}

\begin{figure}[!ht]
\begin{center}
\includegraphics[scale=0.78]{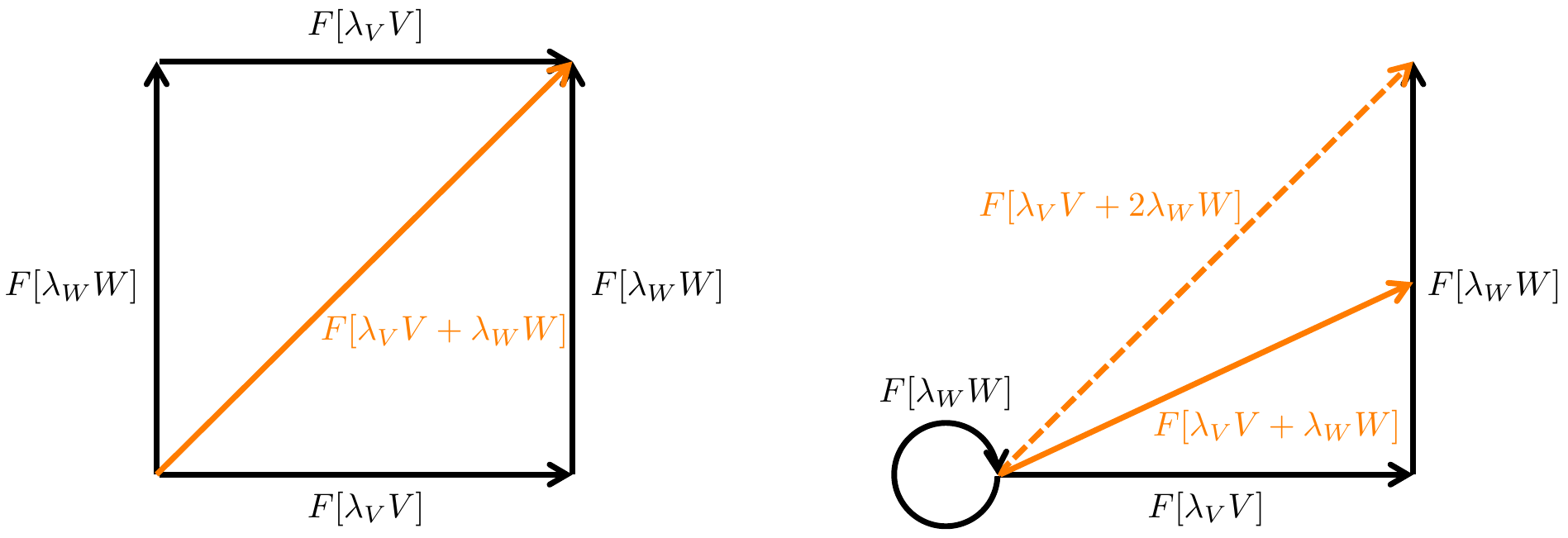}
\caption{{\bf Left:} Graphical representation of \eqref{CouplingSpace2}. Independent of which order we evolve the spectrum of the CFT with $V$ and $W$, we get the same spectrum. The result also agrees by the simultaneous irrelevant deformation by $V$ and $W$ represented by the diagonal orange arrow. equation {\bf Right:} For the special case of $V=J\Tbar,\, W=J\Theta$, the structure of the coupling space is more complicated, as explained in \eqref{CouplingSpace1}. \label{fig:CouplingSpace}}
\end{center}
\end{figure}

To explore some properties of the spectrum, we ask  what happens if we turn on the deformations one after another instead of simultaneously, which gives  \eqref{BrutalFormula} . See Figure~\ref{fig:CouplingSpace} for a sketch of the situation. We will only work with two couplings and turn off all the others. We leave the exploration of more complicated paths in coupling space for future work.

Let us turn on $\lam_{V}$ first, where $V$ is one of the five irrelevant operators that we are studying and $\lam_V$ is the dimensionful version of $g_V$, see \eqref{Dimless}. The spectrum with linear deformations turned on is obtained by setting $g_{\sO}=0,\, (\sO\neq V)$ in \eqref{BrutalFormula}. We can use this result as initial condition for the $\lam_{W},\, (W\neq V)$ flow  equations for the spectrum given in Table~\ref{tab:energyflow}. This is a more complicated initial condition than the conformal initial conditions \eqref{FullFlow}. Nevertheless, the flow is still solvable in a closed form. Setting the background fields to zero at the end, we obtain the spectrum of the theory first deformed by $V$ and then by $W$, which we denote by $F[\lam_W W]F[\lam_V V]\sig_{CFT}$, where $\sig_{CFT}$ is the CFT spectrum and $F[\lam_\sO \sO]$ is a symbolic operator implementing the flow. We find that the only nontrivial flow is obtained for $V=J\Tbar,\, W=J\Theta$ (and similarly for their conjugates), for which
\es{CouplingSpace1}{
F[\lam_{J\Tbar} \, J\Tbar]F[\lam_{J\Theta}\,J\Theta]\sig_{CFT}&=F[\lam_{J\Tbar}\, J\Tbar]\sig_{CFT} \,,\\
F[\lam_{J\Theta}\,J\Theta]F[\lam_{J\Tbar} \,J\Tbar] \sig_{CFT}&=F[\lam_{J\Tbar} \,J\Tbar+2 \lam_{J\Theta}\, J\Theta]\sig_{CFT}\,.
}
The first equation is easy to understand: in a CFT $J\Theta=0$, hence $F[\lam_{J\Theta}] \sig_{CFT}= \sig_{CFT}$. The second equation is the result of a nontrivial computation.  By $F[\lam_V V+\lam_W W]$ we mean the specific flow that led to \eqref{BrutalFormula}, i.e.~we use the common scale $\ell$ as defined in  \eqref{Dimless} and flow with the combined flow equation \eqref{FullFlow}. Note the factor of 2 multiplying $\lam_{J\Theta}$ in the second equation. Because \eqref{BrutalFormula} only depends on $\hat G_{T\Tbar}$, which is a linear combination of $\lam_{J\Tbar}\lam_{J\Theta}$ and $\lam_{T\Tbar}$, we could have written $F[\lam_{J\Tbar} \,J\Tbar+2 \lam_{J\Theta}\, J\Theta]$ equivalently as $F[\lam_{J\Tbar} \,J\Tbar+\#\, J\Theta+\# T\Tbar]$.  For any other pair of operators:
\es{CouplingSpace2}{
F[\lam_W W]F[\lam_V V]\sig_{CFT}&=F[\lam_W W]F[\lam_V V]\sig_{CFT}=F[\lam_V V+\lam_W W]\sig_{CFT}\,.
}

We conclude that the structure of the coupling space is rather simple, as most deformations commute. In particular making two deformation in succession does not lead out of the space of theories that we can reach by the simultaneous deformation by all operators, as given in \eqref{BrutalFormula}.

\subsection{Solving and checking the \texorpdfstring{$J\bar J$}{JJbar} deformation }

Recall that because we were using a dimensionful parameter to control the flow, we did not cover the case of the $J\bar J$ deformation, when solving \eqref{FullFlow}. The solution of the differential equation given in the first row of Table~\ref{tab:energyflow} is a lot simpler than that of \eqref{FullFlow}. The introduction of the same variables as in \eqref{BrutalFormula} is useful, and we get:
\es{JJbarSol}{
\ep_n=E_n \hat L=s+\hat{Q}\hat{\bar Q} \sinh (2\pi \lam_{J\bar J})+(\hat{Q}^2+\hat{\bar Q}^2)\sinh^2 (\pi \lam_{J\bar J})\,.
}
In \cite{Dong:2014tsa} the change in the scaling dimension of certain primary operators was obtained using AdS/CFT and confirmed to second order in perturbation theory. The result in their equation (5.1) for equal left and right levels $k=\tilde k$, and after redefining $q=\sqrt{k\ov2}Q,\, \tilde q=\sqrt{k\ov2}\bar Q,\, h\equiv{2H\ov k\pi}$, reads
\es{JJbarDong}{
\ep_n=\ep_0-{2H\ov 1-H^2}Q\bar Q+{H^2\ov 1-H^2}(Q^2+\bar Q^2)\,.
} 
Because $H$ and $\lam_{J\bar J}$ are dimensionless, and the space of theories is one dimensional, different definitions can give different parametrizations of the same line. Indeed setting
\es{lamH}{
\lam_{J\bar J}=-{1\ov 2\pi}{\rm arcsinh} \le({2H\ov 1-H^2}\ri)
} 
in \eqref{JJbarSol} and setting the background fields to zero produces \eqref{JJbarDong}. This is another nice check of the validity of our formalism. 

\subsection{A check from string theory}\label{sec:String}

In this section we use the string construction of \cite{Giveon:2017nie,Giveon:2017myj,Chakraborty:2018vja} to test a special case of the energy formula  \eqref{SpectrumGuess}, where the background fields are set to zero $\sA=\bar \sA=b=0$ and the $J\Theta$ and $\bar J\bar \Theta$ deformations are turned off. We obtain a precise match.

We now give a lightning review of the argument of  \cite{Chakraborty:2018vja}, skipping over many important details. Let us consider Type II superstrings on the background (massless BTZ)$\times S^1\times \sN$. Vertex operators of the worldsheet theory dual to certain Ramond sector states of the dual CFT$_2$ were constructed in \cite{Chakraborty:2018vja}, whose explicit form we will not need.  The construction uses separate primaries both in the (massless BTZ)$\times S^1$ and the $\sN$ CFTs. The Virasoro constraint imposes:
\es{VirConst}{
0&=\Delta_1+\Delta_2-\ha\,,\\
0&=\bar\Delta_1+\bar\Delta_2-\ha\,,
}
where $\Delta_{1,2}$ are the left scaling dimensions in the (massless BTZ)$\times S^1$ and $\sN$ CFTs respectively. To simplify formulas, we restrict to the winding number 1 sector of the theory. The scaling dimensions in the (massless BTZ)$\times S^1$ CFT are given by:
\es{Delta1Form}{
\Delta_1&=-E_L+{Q^2\ov 8\pi^2}-{j(j+1)\ov k}\,,\\
\bar\Delta_1&=-E_R+{\bar Q^2\ov 8\pi^2}-{j(j+1)\ov k}\,,\\
E_{L,R}&\equiv\ha (E\pm P)\,,
}
where $j$ is a quantum number related to radial motion, $k=\le(L_\text{AdS}\ov \ell_s\ri)^2$, and we set $R=1, \, q_L={Q\ov 2\pi},\, q_R={\bar Q\ov 2\pi}$ in the formulas of \cite{Chakraborty:2018vja}.\footnote{A check on these normalization factors is that if we set $r=\sqrt{2}$ in \eqref{ChargesEval}, we get the same spectrum of scaling dimensions, as in \cite{Chakraborty:2018vja}.}

The (massless BTZ)$\times S^1$ CFT is an $SL(2,\R)_k \times U(1)$ WZW model, and hence has interesting exactly marginal $J\bar{J}$ deformations. It was argued in \cite{Giveon:2017nie,Chakraborty:2018vja} that a deformation that to linear order agrees with the $J_{SL}^-\bar{J}_{SL}^-$ deformation in the terminology of this paper is related to a single trace version of the $T\Tbar$ deformation of the dual CFT, while the $J_{U(1)}\bar{J}_{SL}^-$ and $J_{SL}^-\bar J_{U(1)}$ deformation is related to the $J\Tbar$ and $\bar J T$ deformations \cite{Chakraborty:2018vja}. The arguments are complicated and rely on some conjectures about the dual CFT; the proposal is that the string states created by the vertex operators discussed above should evolve in the same way under the single trace deformation version of the irrelevant deformations as they would under their double trace versions to which our field theory treatment applies.

Under the $J\bar{J}$ deformations of the (massless BTZ)$\times S^1$ CFT the scaling dimensions in the  $\sN$ CFT do not change, while the change of $\Delta_1$ and $\bar \Delta_1$ can be determined by combining formulas (5.23) and (5.34) from \cite{Chakraborty:2018vja}. Adapting their equations to our notation, which includes introducing the coupling constants $g_{ J \Tbar},\,g_{\bar J T},\, g_{T\Tbar}$ with appropriate numerical prefactors, gives
\es{Delta1Def}{
\Delta_1&=-E_L+{Q^2\ov 8\pi^2}-g_{\bar J T}\mu \bar Q E_L+{2\pi^2 g_{\bar J T}^2\mu^2 E_L^2}-g_{J\Tbar}\mu Q E_R+2\pi^2 g_{J\Tbar}^2\mu^2 E_R^2+4\pi g_{T\Tbar}\mu^2E_L E_R-{j(j+1)\ov k}\,,\\
\bar\Delta_1&=-E_R+{\bar Q^2\ov 8\pi^2}-g_{\bar J T}\mu \bar Q E_L+{2\pi^2 g_{\bar J T}^2\mu^2 E_L^2}-g_{J\Tbar}\mu Q E_R+2\pi^2 g_{J\Tbar}^2\mu^2 E_R^2+4\pi g_{T\Tbar}\mu^2E_L E_R-{j(j+1)\ov k}\,.
}
Subtracting the two equations leads to a $\mu$ independent result, which expresses that the spin of the vertex operator is quantized. Henceforth, we drop the second equation. We remark that the field theory explanation of why we can simply add together the contributions of different deforming operators is that these deformations commute in the sense explained in Section~\ref{sec:Couplingspace}. We also note that linearizing in the couplings $g_{J\Tbar},\,g_{\bar J T},\, g_{T\Tbar}$, we get
\es{EnergyChange}{
\de E=\de(\Delta_1+\bar\Delta_1)=-2g_{J\Tbar} Q (\mu E_R)-2g_{\bar J T} \bar Q (\mu E_L)+4\pi g_{T\Tbar}(\mu E_L)(\mu E_R)\,,
}
which to linear order agrees with the spectrum of the  $J\bar J$-deformed theory \eqref{JJbarSol}, if we identify the charges of $J_{U(1)},\, \bar J_{U(1)},\, J_{SL}^-,\, \bar{J}_{SL}^-$ with $Q,\, \bar Q,\, (\mu E_L),\, (\mu E_R)$, and 
$g_{J\Tbar},\, g_{\bar J T} ,\, g_{T\Tbar}$ with (up to constant factors) the coupling $\lam_{J \bar J}$ of the $J\bar J$ operators $J_{U(1)}\bar{J}_{SL}^-,\, J_{SL}^-\bar J_{U(1)},\, J_{SL}^- \bar{J}_{SL}^-$. To higher orders the agreement does not hold, demonstrating that the worldsheet CFT is not precisely a $J\bar J$ deformation of the (massless BTZ)$\times S^1$ CFT in the terminology of this paper.

To get the evolution of energies in the boundary theory, we put $(0)$ superscripts on the quantities in the undeformed theory given in \eqref{Delta1Form}. The Virasoro constraint \eqref{VirConst} implies that we have to equate $\Delta_1$ (and $\bar \Delta_1$) in \eqref{Delta1Form} and \eqref{Delta1Def}, which then implies:
\es{FinalEq}{
-E^{(0)}_L+{Q^2\ov 8\pi^2}&=-E_L+{Q^2\ov 8\pi^2}-g_{\bar J T}\mu \bar Q E_L+{2\pi^2 g_{\bar J T}^2\mu^2 E_L^2}-g_{J\Tbar}\mu Q E_R+2\pi^2 g_{J\Tbar}^2\mu^2 E_R^2+4\pi g_{T\Tbar}\mu^2E_L E_R\,.
}
This equation can be brought to the form:
\es{Brutal2}{
0=(2\pi)^2A(E-E^{(0)})^2+2\pi B (E-E^{(0)})+C\,,
}
which equivalent to \eqref{BrutalFormula} with all the background fields turned off, $\sA=\bar \sA=b=0,\, \hat Q=Q,\,\hat{\bar Q}=\bar Q,\, G_\sO=g_\sO$, and with the $J\Theta$ and $\bar J\bar \Theta$ deformations turned off, $g_{J\Theta}=g_{\bar J\bar \Theta}=0$. The factors of $2\pi$ account for the fact that we set $L=2\pi$ by setting $R=1$, hence $\ep=2\pi E$. (Also $p=2\pi P$.)

\section{Conclusions and outlook}\label{sec:Conclusions}

In this paper, we presented detailed arguments for the proposal of the energy spectrum of CFTs deformed simultaneously by  $J\Tbar,\, J \Theta,\, \bar J T,\, \bar J \bar\Theta,\, T\Tbar$ and also by the linear deformations $J_x,\, \bar J_x,\,T_{tx}$ which are equivalent to turning on the background fields $a,\, \bar a,\, b$ in \eqref{BrutalFormula}.
 Note that deforming by the time component of conserved currents would be trivial: since the corresponding charges commute with the Hamiltonian,  the eigenvalues $Q,\, \bar Q,\, E,\, P$ would simply add under such deformations.

We have arrived at this spectrum following the strategy outlined in Section~\ref{sec:strategy}, see also Figure~\ref{fig:Strategy}. We implemented the first step of the strategy, rigorously determining the initial conditions for the flow equation for the Hamiltonian. We then used the example of the classical free scalar to conjecture a universal equation valid at an arbitrary point in coupling space in \eqref{JTbarFinal} and Table~\ref{tab:ddlam}. We performed two checks of this equation in Section~\ref{sec:checks}: we checked a special case of the equation in a more general classical field theory setting in Section~\ref{sec:CLcheck}, and a quantum mechanical check in low order perturbation theory in  Section~\ref{sec:PertCheck}. It would be interesting to go to higher orders in perturbation theory. Ultimately, in the future we would like to find a nonperturbative quantum derivation of these equations. From these universal equations, the flow equation for the energy follows using the factorization property of the special composite operators discussed in this paper; this is again a fully rigorous step. We then solved this equation using the initial conditions to obtain the spectrum \eqref{BrutalFormula}. That we have obtained elegant solvable equations for the spectrum, whose solution reproduces previously known special cases gives us confidence that this is the full quantum answer. We solved the $J\bar J$ deformation with the same methods reproducing the spectrum found  using conformal perturbation theory and AdS/CFT in \cite{Dong:2014tsa}. We also did a new AdS/CFT computation in Section~\ref{sec:String} of the spectrum of the  $J\Tbar,\, \bar J T,\, T\Tbar$ deformed theory with the background fields turned off, and this confirmed the spectrum \eqref{BrutalFormula} in a special case. We provided evidence in Section~\ref{sec:Couplingspace}, that turning on the couplings of the irrelevant operators in different order (instead of simultaneously) will not lead out of the space of theories we solved. 

This work leaves many interesting directions open for future investigation. It would be interesting to obtain the spectrum of non-conformal theories deformed by irrelevant operators. While the universal operator equations and hence the flow equations for the spectrum apply, we do not know how to obtain the initial conditions. An exception is provided by Gaussian theories, the massive complex boson and fermion, where linear deformations preserve their Gaussian nature and hence solvability. We leave their solution to future work.
The universal equations for the Hamiltonian can be solved for the classical Hamiltonian (and Lagrangian). We have only discussed this in Appendix~\ref{app:recover}, since we do not know how to quantize these theories starting from the Lagrangian. Since, we do know how to understand these theories using flow equations, this could give insight into how to treat such exotic theories that include the Nambu-Goto string (in static gauge) \cite{Dubovsky:2012wk,Cavaglia:2016oda}. 
We would also like to understand, how to turn on the background fields in the AdS/CFT setup analyzed in Section~\ref{sec:String}. 
It would be interesting to analyze the torus partition function of this class of theories, and fascinating to  
prove uniqueness results similar to those found in \cite{Aharony:2018bad,Aharony:2018ics}. 

The most interesting extension of this work would be to understand deformations by quadratic composite operators built from the higher spin KdV currents. Understanding these would presumably lead to qualitatively new UV behaviors. In Section~\ref{sec:KdV} we attempted to obtain the initial conditions for this flow, and explained why our approach does not apply straightforwardly. Despite this failure, it would be interesting to understand, if there exists a universal operator equation governing the Hamiltonian at an arbitrary point in coupling space. Since the background field for the spin $s$ current is irrelevant in this case, $[\al_s]=2-s$, we expect a proliferation of terms. A first step in this direction would be to work out the case of the classical free scalar. We also note that we take a step in a tanglential direction in a future publication, where we compute the spectrum of KdV conserved charges in the $T\Tbar$ flow. 

\section*{Acknowledgements}

We thank Costas Bachas, Shouvik Datta, Guido Festuccia, Yunfeng Jiang, David Kutasov, Philippe LeFloch, Stefano Negro, Ilia Smilga, Tin Sulejmanpasic, Jan Troost, Herman Verlinde, Yifan Wang for useful discussions and Amit Giveon for coordinating the arXiv submission of \cite{note} with our paper. MM is supported by the Simons Center for Geometry and Physics.

\appendix

\section{Conventions}\label{app:conv}

We use the following conventions. The complex coordinates in Euclidean space are defined by:
\es{zzb}{
z=x+i t\,, \qquad \bar z=x-it\,.
}
The cylinder $S^1\times R$ has circumference $L$, hence $z\sim z+L$.
The stress tensor in a relativistic theory is defined by
\es{StressDef}{
T_{\mu\nu}\equiv{2\ov \sqrt g}{\de S\ov \de g^{\mu\nu}}\,.
}
We will be dealing with theories that cannot be easily coupled to gravity, as they do not have Lorentz invariance, and for these we will be using the Noether stress tensor that obeys:
\es{StressCons}{
0&=\p^\nu T_{\mu\nu}\,,
}
with the corresponding (Hermitian) conserved quantities:
\es{ConsQuant}{
H=-\int_0^L dx \ T_{tt}\,, \qquad P=-i\int_0^L dx \ T_{xt}\,.
}
They generate the spacetime translations:
\es{Trans}{
[H,\sO]=\p_t \sO\,, \qquad [P,\sO]=i\p_x \sO\,.
}
One useful example to keep in mind is that for a scalar theory whose Lagrangian $\sL(\phi,\p_\mu\phi)$ does not contain higher derivatives (but may contain arbitrary powers of single derivatives), we have
\es{StressDef2}{
T_{\mu\nu}&\equiv{\p \sL\ov \p (\p^\nu \phi)}\p_\mu\phi-\de_{\mu\nu} \sL\,,
}
which obeys \eqref{StressCons} by the equations of motion.

In CFT the components of the stress tensor are usually denoted by
\es{StressComp}{
  T &\equiv -2\pi T_{\z\z}\,, \qquad \Theta\equiv +2\pi T_{\z\zbar}\,, \qquad \Thetabar \equiv + 2\pi T_{\zbar \z} \,, \qquad \Tbar \equiv -2\pi T_{\zbar\zbar} \,, 
  }
  and the conservation equation written as
  \es{StressCompCons}{
  \bar\p T&=\p \Theta\,, \qquad \p \Tbar=\p \Thetabar\,.
  }
 For convenience, we also  write down the components of the stress tensor in $(x,t)$ coordinates:
\es{StressComp2}{  
   T_{\rmt\rmt}&={1\ov 2\pi} \le(T + \Theta + \Thetabar + \Tbar\ri)\,,\qquad T_{\rmt\rmx}=-{i\ov 2\pi}\le(T-\Theta+\Thetabar-\Tbar \ri)  \,,\\
  T_{\rmx\rmt}&=-{i\ov 2\pi}\le( T+\Theta-\Thetabar-\Tbar \ri) \,, \qquad T_{\rmx\rmx}=-{1\ov 2\pi}\le( T - \Theta - \Thetabar + \Tbar \ri)   \,.
}

For a conserved current corresponding to an internal symmetry, we have
\es{ConsCurr}{
0&=\p^\mu J_\mu\\
Q&=\int_0^L dx \ J_t(x)
}
From the perspective of Euclidean field theory a natural formula would instead be $-i\int_0^L dx \ J_t(x)$, because we want $Q$ to be Hermitian, and to Wick-rotate an operator with spin, we conventionally add factors of $i$
to the time components. However, to conform with CFT convention, we chose to omit the $i$ from \eqref{ConsCurr}. If we have only one current a natural choice for normalization is that the charge is an integer. However, as we recall on the  example of the free compact boson in Appendix~\ref{app:scalar}, this normalization is not always natural. In CFT, it is customary to use the notation $J\equiv J_z,\, \bar J \equiv \bar J_{\bar z}$.

\section{Free compact boson}\label{app:scalar}

We take the free massless scalar Lagrangian to be:
\es{FreeScaLag}{
\sL={1\ov 8\pi}(\p_\mu\phi)^2\,.
}
The equation of motion is:
\es{EoM}{
0&=(\p_t^2+\p_x^2)\phi\,.
}
The propagator is
\es{Propagator}{
\<\phi(z)\phi(0)\>=-\log\abs{z}^2\,.
}
The currents and the stress tensor (on the plane) are given by:
\es{Quants}{
J&=i\p\phi\,, \qquad \bar J=-i\bar\p\phi\,,\\
T&=-\frac12\nop{(\p\phi)^2}=\frac12\nop{J^2}\,, \qquad \bar T=-\frac12\nop{(\bar\p\phi)^2}=\frac12\nop{\bar J^2}\,.
}

The canonical momentum and Hamiltonian densities are defined by
\es{CanMom}{
\Pi&\equiv i {\p L\ov \p(\p_t\phi)}={i\ov 4\pi} \p_t \phi \\
\sH&\equiv i \Pi \p_t \phi+\sL=2\pi \Pi^2+{1\ov 8\pi}(\p_x\phi)^2\,.
}
The canonical commutation relations are $\le[\Pi(x),\phi(y)\ri]=-i\de(x-y)$. It is easy to check that $\sH=-{1\ov 2\pi} \le(T+ \Tbar\ri)$, consistent with \eqref{ConsQuant} and \eqref{StressComp2}. Hamilton's equations are:
\es{HamEoM}{
\p_t \Pi&=i{\de H\ov \de \phi}=-i\p_x \le({\p \sH\ov \p(\p_x\phi)}\ri)\,,\\
\p_t \phi&=-i{\de H\ov \de \Pi}=-i\le({\p \sH\ov \p\Pi}\ri)\,,\\
}
which are consistent with \eqref{EoM}. Since the Hamiltonian is the same as in usual Lorentzian quantum mechanics, \eqref{HamEoM} are just the usual Lorentzian Hamilton's equations with the replacement ${\p\ov \p t_L}=i{\p\ov \p t}$.

We take the boson to be compact with radius $r$:
\es{Compact}{
\phi\sim\phi+2\pi r\,.
}
The mode expansion of the scalar and the currents on the cylinder is:
\es{phiMode}{
\phi&=\phi_0+{4\pi n\ov r L}(-it)-{2\pi r w\ov L} x+\text{(oscillating terms)}\\
J&=i\p \phi =-{i\ov2}\le({4\pi n\ov r L}+{2\pi r w\ov L}\ri)+\text{(oscillating terms)}\\
\bar J&=-i\bar \p \phi ={i\ov2}\le({4\pi n\ov r L}-{2\pi r w\ov L}\ri)+\text{(oscillating terms)}\,.
}
Then the charges using \eqref{ConsCurr} are:
\es{ChargesEval}{
Q={2\pi n\ov r }+{\pi r w }\,, \qquad \bar Q={2\pi n\ov r }-{\pi r w}\,.
}
The charges are quantized, but are not integers. In fact, for irrational $r$, there does not exist a normalization in which they would be integer valued.\footnote{Except at the self-dual radius $r=\sqrt{2}$. }

At $\lam=0$, solving the differential equations \eqref{JxJbarx} and \eqref{bguess} with the expressions for the conserved current components given in \eqref{ConsCurrComps} gives:
\es{DeformedHam}{
\sH&={ 2\pi \le(\Pi+{a-\bar a\ov 2}\ri)^2+{1\ov 8\pi}\le(\p_x\phi-2\pi (a+\bar a)\ri)^2-b \le(\Pi+{a-\bar a\ov 2}\ri) \le(\p_x\phi-2\pi (a+\bar a)\ri)\ov 1-b^2}\,.
}
The shifts of $\Pi$ and $\p_x\phi$ are explained by the comment around \eqref{JxJbarx}.

\section{The Hamiltonian and Lagrangian of the deformed free scalar}\label{app:recover}

We have not determined the closed form solution of the flow equations for the Hamiltonian density, \eqref{JTbarFinal} and Table~\ref{tab:ddlam}. They can be recovered from the solution of the spectrum \eqref{BrutalFormula} using the following very simple recipe. In $\ep_n(\ep_0,p,Q,\bar Q)$ as a function of the four initial conditions, we have to make the replacements:
\es{Hform}{
\sH={1\ov (1-b^2)L^2}\,\ep_n\le(L^2\le(2\pi \Pi^2+{1\ov 8\pi}(\p_x\phi)^2 \ri),\, -L^2\,\Pi \p_x \phi, \,- \frac{L}{2}(\del_\rmx\phi-4\pi \Pi),\, -  \frac{L}{2}(\del_\rmx\phi+4\pi \Pi) \ri)\,.
}
 The intuitive way to obtain this formula is to take a simple classical phase space configuration, for which
\es{ClassConf}{
\p_x\phi&=-{2\pi r w\ov L}\,, \qquad \Pi={ n\ov r L}\,, \\
\ep_0&=2\pi \le({n^2\ov r^2}+{r^2 w^2\ov 4}\ri)\,, \qquad p=2\pi n w\,, \qquad Q={2\pi n\ov r }+{\pi r w }\,, \qquad \bar Q={2\pi n\ov r }-{\pi r w}\,.
}
In the undeformed case, the appropriate field configuration of $\phi$ is given in \eqref{phiMode} (with the oscillating terms set to zero), but after deformation $\Pi\neq {i\ov 4\pi} \p_t \phi$. Plugging these into $\sH$ defined by \eqref{Hform} and multiplying by $L$ to perform the trivial integral over $x$, we recover the spectrum \eqref{BrutalFormula}. This is not quite a proof of \eqref{Hform}, since for the special configurations in \eqref{ClassConf} $\ep_0$ and $p$ are determined by $Q,\, \bar Q$. Nevertheless, it can be checked explicitly that  $\sH$ resulting from \eqref{Hform} indeed solves the appropriate equations. We note, that \eqref{DeformedHam} can indeed be recovered by using \eqref{Hform}, with $\ep_n$ replaced by the initial condition $s$ form \eqref{BrutalFormula}. Finally, the Lagrangian for the deformed theories can be obtained by Legendre transformation,
\es{LegendreTF}{
{\p\sH\ov \p \Pi}&=i \p_t\phi\,,\\
\sL&=- i\Pi\,\p_t\phi+\sH\,.
}

Conversely, given a Lagrangian of a shift symmetric scalar field $\sL(\p\phi,\bar\p\phi)$, we can conjecture its energy spectrum by first going to the Hamiltonian $\sH(\p_x\phi,\Pi)$, plugging in the first line of \eqref{ClassConf}, and expressing $n,\, w$ with the possible initial conditions, $\ep_0,\, p,\, Q,\, \bar Q$. As already mentioned above, since there are four initial conditions and only the two $n,\, w$ in the output of this procedure, obtaining the right spectrum this way requires some guesswork. In the case of the $T\Tbar$ deformation (with background fields turned off), the spectrum  cannot depend on $Q,\, \bar Q$ and one obtains the correct spectrum from the Lagrangian
\es{LTTbar}{
\sL_{T\Tbar}&={1\ov 2\pi^2\ell^2}\le(1-\sqrt{1-{\pi\ell^2\ov 2}(\p_\mu\phi)^2}\ri)\,
}
as was shown in \cite{Kraus:2018xrn} using the method described here.

\section{Comments on the \texorpdfstring{$J\Tbar$}{JTbar} deformation}\label{app:JTbar}

At first sight, it may seem that our definition of the $J\Tbar$ deformation and the one in the literature is different. We have been working with a non-holomorphic current that had quantized charge that does not change under the deformation, while in the literature the current $J$ is holomorphic and its charge depends on the scale $\ell$ \cite{Chakraborty:2018vja}. It turns out that the two definitions of the theories are equivalent, as we explain below. 

Let us determine the explicit form of $\sH$ for the case of the $J\Tbar$ deformation with the background fields turned off. We can use the simple explicit formula given in \eqref{JTbarSol} instead of \eqref{BrutalFormula}. Plugging into \eqref{Hform} we obtain:
\es{HJTbar}{
\sH_{J\Tbar}&={1-{\pi\ell\ov 2}(\del_\rmx\phi-4\pi \Pi)-\pi^3\ell^2\Pi \p_x \phi -\sqrt{\le(1+4\pi^2 \ell \Pi\ri)\le(1-\pi \ell \p_x \phi\ri)}\ov \pi^3 \ell^2}\,,\\
\sL_{J\Tbar}&={1\ov 2\pi}\p\phi{\bar \p\phi\ov 1-\pi\ell \bar \p\phi}\,.
}
The latter Lagrangian was obtained in \cite{Guica:2017lia,Chakraborty:2018vja}. It can be checked that the current
\es{Jhat}{
\hat J_\mu\equiv J_\mu-2\pi^2 i \ell\,  T_{\zbar \mu}
}
is holomorphic, once we plug in $\sH$ from \eqref{HJTbar} in the expression of the conserved current components \eqref{ConsCurrComps}. So the current used in the literature is just the linear combination of currents defined in our current formalism. The ambiguity of combining currents was discussed in Section~\ref{sec:Ambiguity}.

Finally, we show that the operators $J\Tbar$ and $\hat J\Tbar$ are the same, hence the deformed theories are equivalent. Writing out the definition \eqref{QuadraticOps}, we get
\es{2JTbars}{
\hat J\Tbar &= 2\pi  i\hat J_{[t|}T_{\zbar|x]}=J\Tbar-4\pi^3\ell\, T_{\zbar [t}T_{\zbar|x]}\\
&=J\Tbar\,,
}
which is just the manifestation of the simple fact that the quadratic composite operator built from the same current is identically zero, $\sO\equiv\ep^{\mu\nu}J_\mu J_\nu=0$.

\section{Quantum perturbation theory formulas}\label{app:qupertformulas}

We collect here some formulas relevant to Section~\ref{sec:PertCheck}.  Local operators are expanded in modes as
\begin{equation}
  \Ocal(x) = \biggl(\frac{2\pi}{L}\biggr)^\Delta \sum_{n=-\infty}^\infty e^{2\pi inx/L} \Ocal_n
\end{equation}
where $\Ocal_n$ are dimensionless and $\Delta$ is the dimension of~$\Ocal(x)$.
We denote the CFT (anti-)holomorphic current and stress tensor by $J$, $T$, $\Tbar$ (signs and factors of $i$ chosen to match standard 2d CFT literature):
\begin{equation}\label{CFTTTbarJinmodes}
  \begin{aligned}
    T & = - \biggl(\frac{2\pi}{L}\biggr)^2 \sum_{n=-\infty}^\infty e^{2\pi inx/L} \, \ell_n , \\
    \Tbar & = - \biggl(\frac{2\pi}{L}\biggr)^2 \sum_{n=-\infty}^\infty e^{2\pi inx/L} \, \bar{\ell}_n , \\
    J & = - i \frac{2\pi}{L} \sum_{n=-\infty}^\infty e^{2\pi inx/L} \, j_n ,
  \end{aligned}
\end{equation}
and we recall $T_{tt}^{\text{CFT}} = - T_{xx}^{\text{CFT}} = (T + \Tbar)/(2\pi)$ and $T_{xt}^{\text{CFT}} = T_{tx}^{\text{CFT}} = (T - \Tbar)/(2\pi i)$ and $J_t^{\text{CFT}} = i J_x^{\text{CFT}} = i J$.

We shifted $\ell_m\equiv L_m - \delta_{m,0}\,c/24$ and $\bar{\ell}_m\equiv \overline{L}_m - \delta_{m,0}\,c/24$ compared to the usual Virasoro algebra, and these modes have non-zero commutators
\begin{equation}\label{appEVirasoroKacMoody}
  \begin{aligned}
    [\ell_m , \ell_n] & = (m-n)\ell_{m+n} + \frac{c}{12} m^3 \delta_{m+n,0} ,
    & [\ell_m , j_n] & = -n j_{m+n} , \\
    [ \bar{\ell}_m , \bar{\ell}_n ] & = (m-n)\bar{\ell}_{m+n} + \frac{c}{12} m^3 \delta_{m+n,0} ,
    & [j_m,j_n] & = m \delta_{m+n} .
  \end{aligned}
\end{equation}

After deformation by $J\Tbar$ we find spectrum-generating operators (we do not display order $\lambda^2$ terms in $\Lambda_k$ and $\overline{\Lambda}_k$ because they are too long)
\begin{equation}\label{explicitLambdaetc}
  \begin{aligned}
    \Upsilon_k & = j_k + \delta_{k\neq 0} \frac{2\pi^2 i \lambda}{(1+b)L} \bar{\ell}_{-k}
    + \delta_{k\neq 0} \biggl( \frac{2\pi^2 i \lambda}{(1+b) L}\biggr)^2 \biggl(
    (j_0 + L a) \bar{\ell}_{-k} + \frac{1}{2} \sum_{m\neq 0} \frac{m-k}{m} j_m \bar{\ell}_{m-k}
    \biggr) + O(\lambda^3), \\
    \Lambda_k & = \ell_k + \frac{2\pi^2 i \lambda}{(1+b)L} \sum_{m\neq 0} j_{k+m} \bar{\ell}_m + O(\lambda^2), \\
    \overline{\Lambda}_k & = \bar{\ell}_k + \frac{2\pi^2 i \lambda}{(1+b)L} \biggl( \frac{c}{12} k^2 j_{-k} + \sum_{m\neq 0} \frac{m-k}{m} j_m \bar{\ell}_{k+m} \biggr) + O(\lambda^2) .
  \end{aligned}
\end{equation}
Note that $\Upsilon_0 = j_0$ and $\Lambda_0-\overline{\Lambda}_0 = \ell_0 - \bar{\ell}_0$ as expected.
Calculating commutators confirms that these operators obey the same algebra~\eqref{sameVirasoro} as the original modes.

For the $J\Tbar$ deformation we find
\begin{equation}\label{JTbarHamiltonianAsLambda}
  \begin{aligned}
    H & = \frac{2\pi}{L} \biggl(\frac{\Lambda_0}{1-b} + \frac{\overline{\Lambda}_0}{1+b} + \frac{a L \Upsilon_0}{1-b} + \frac{a^2 L^2}{2(1-b)} \biggr)
    + 2 \pi i \lambda \biggl(\frac{2\pi}{L}\biggr)^2 \frac{(\Upsilon_0 + a L) \overline{\Lambda}_0}{(1-b)(1+b)^2} \\
    & \quad + \frac{(2\pi i\lambda)^2}{2} \biggl(\frac{2\pi}{L}\biggr)^3
    \frac{(\Upsilon_0 + a L)^2\overline{\Lambda}_0 + \overline{\Lambda}_0^2/2}{(1-b)(1+b)^3}
    + O(\lambda^3)
  \end{aligned}
\end{equation}

\clearpage

\bibliography{TTbarrefs}

\providecommand{\href}[2]{#2}\begingroup\raggedright\begin{thebibliography}{10}

\bibitem{Zamolodchikov:2004ce}
A.~B. Zamolodchikov, \emph{{Expectation value of composite field T anti-T in
  two-dimensional quantum field theory}},
  \href{https://arxiv.org/abs/hep-th/0401146}{{\ttfamily hep-th/0401146}}.

\bibitem{Smirnov:2016lqw}
F.~A. Smirnov and A.~B. Zamolodchikov, \emph{{On space of integrable quantum
  field theories}},
  \href{https://doi.org/10.1016/j.nuclphysb.2016.12.014}{\emph{Nucl. Phys.}
  {\bfseries B915} (2017) 363}
  [\href{https://arxiv.org/abs/1608.05499}{{\ttfamily 1608.05499}}].

\bibitem{Cavaglia:2016oda}
A.~Cavagli{\`a}, S.~Negro, I.~M. Sz{\'e}cs{\'e}nyi and R.~Tateo, \emph{{$T
  \bar{T}$-deformed 2D Quantum Field Theories}},
  \href{https://doi.org/10.1007/JHEP10(2016)112}{\emph{JHEP} {\bfseries 10}
  (2016) 112} [\href{https://arxiv.org/abs/1608.05534}{{\ttfamily
  1608.05534}}].

\bibitem{Cardy:2018jho}
J.~Cardy, \emph{{$T\overline T$ deformations of non-Lorentz invariant field
  theories}},  \href{https://arxiv.org/abs/1809.07849}{{\ttfamily 1809.07849}}.

\bibitem{Chakraborty:2018vja}
S.~Chakraborty, A.~Giveon and D.~Kutasov, \emph{{$ J\overline{T} $ deformed
  CFT$_{2}$ and string theory}},
  \href{https://doi.org/10.1007/JHEP10(2018)057}{\emph{JHEP} {\bfseries 10}
  (2018) 057} [\href{https://arxiv.org/abs/1806.09667}{{\ttfamily
  1806.09667}}].

\bibitem{Guica:2017lia}
M.~Guica, \emph{{An integrable Lorentz-breaking deformation of two-dimensional
  CFTs}}, \href{https://doi.org/10.21468/SciPostPhys.5.5.048}{\emph{SciPost
  Phys.} {\bfseries 5} (2018) 048}
  [\href{https://arxiv.org/abs/1710.08415}{{\ttfamily 1710.08415}}].

\bibitem{Nakayama:2018ujt}
Y.~Nakayama, \emph{{Very Special $T\bar{J}$ deformed CFT}},
  \href{https://arxiv.org/abs/1811.02173}{{\ttfamily 1811.02173}}.

\bibitem{Dubovsky:2012wk}
S.~Dubovsky, R.~Flauger and V.~Gorbenko, \emph{{Solving the Simplest Theory of
  Quantum Gravity}}, \href{https://doi.org/10.1007/JHEP09(2012)133}{\emph{JHEP}
  {\bfseries 09} (2012) 133} [\href{https://arxiv.org/abs/1205.6805}{{\ttfamily
  1205.6805}}].

\bibitem{Dubovsky:2013ira}
S.~Dubovsky, V.~Gorbenko and M.~Mirbabayi, \emph{{Natural Tuning: Towards A
  Proof of Concept}},
  \href{https://doi.org/10.1007/JHEP09(2013)045}{\emph{JHEP} {\bfseries 09}
  (2013) 045} [\href{https://arxiv.org/abs/1305.6939}{{\ttfamily 1305.6939}}].

\bibitem{Cooper:2013ffa}
P.~Cooper, S.~Dubovsky and A.~Mohsen, \emph{{Ultraviolet complete
  Lorentz-invariant theory with superluminal signal propagation}},
  \href{https://doi.org/10.1103/PhysRevD.89.084044}{\emph{Phys. Rev.}
  {\bfseries D89} (2014) 084044}
  [\href{https://arxiv.org/abs/1312.2021}{{\ttfamily 1312.2021}}].

\bibitem{Dubovsky:2017cnj}
S.~Dubovsky, V.~Gorbenko and M.~Mirbabayi, \emph{{Asymptotic fragility, near
  AdS$_{2}$ holography and $ T\overline{T} $}},
  \href{https://doi.org/10.1007/JHEP09(2017)136}{\emph{JHEP} {\bfseries 09}
  (2017) 136} [\href{https://arxiv.org/abs/1706.06604}{{\ttfamily
  1706.06604}}].

\bibitem{Dubovsky:2018bmo}
S.~Dubovsky, V.~Gorbenko and G.~Hern{\'a}ndez-Chifflet, \emph{{$ T\overline{T}
  $ partition function from topological gravity}},
  \href{https://doi.org/10.1007/JHEP09(2018)158}{\emph{JHEP} {\bfseries 09}
  (2018) 158} [\href{https://arxiv.org/abs/1805.07386}{{\ttfamily
  1805.07386}}].

\bibitem{Chen:2018keo}
C.~Chen, P.~Conkey, S.~Dubovsky and G.~Hern{\'a}ndez-Chifflet,
  \emph{{Undressing Confining Flux Tubes with $T\bar T$}},
  \href{https://doi.org/10.1103/PhysRevD.98.114024}{\emph{Phys. Rev.}
  {\bfseries D98} (2018) 114024}
  [\href{https://arxiv.org/abs/1808.01339}{{\ttfamily 1808.01339}}].

\bibitem{McGough:2016lol}
L.~McGough, M.~Mezei and H.~Verlinde, \emph{{Moving the CFT into the bulk with
  $ T\overline{T} $}},
  \href{https://doi.org/10.1007/JHEP04(2018)010}{\emph{JHEP} {\bfseries 04}
  (2018) 010} [\href{https://arxiv.org/abs/1611.03470}{{\ttfamily
  1611.03470}}].

\bibitem{Giveon:2017nie}
A.~Giveon, N.~Itzhaki and D.~Kutasov, \emph{{$ \mathrm{T}\overline{\mathrm{T}}
  $ and LST}}, \href{https://doi.org/10.1007/JHEP07(2017)122}{\emph{JHEP}
  {\bfseries 07} (2017) 122}
  [\href{https://arxiv.org/abs/1701.05576}{{\ttfamily 1701.05576}}].

\bibitem{Cardy:2018sdv}
J.~Cardy, \emph{{The $ T\overline{T} $ deformation of quantum field theory as
  random geometry}}, \href{https://doi.org/10.1007/JHEP10(2018)186}{\emph{JHEP}
  {\bfseries 10} (2018) 186}
  [\href{https://arxiv.org/abs/1801.06895}{{\ttfamily 1801.06895}}].

\bibitem{Datta:2018thy}
S.~Datta and Y.~Jiang, \emph{{$T\bar{T}$ deformed partition functions}},
  \href{https://doi.org/10.1007/JHEP08(2018)106}{\emph{JHEP} {\bfseries 08}
  (2018) 106} [\href{https://arxiv.org/abs/1806.07426}{{\ttfamily
  1806.07426}}].

\bibitem{Aharony:2018bad}
O.~Aharony, S.~Datta, A.~Giveon, Y.~Jiang and D.~Kutasov, \emph{{Modular
  invariance and uniqueness of $T\bar{T}$ deformed CFT}},
  \href{https://doi.org/10.1007/JHEP01(2019)086}{\emph{JHEP} {\bfseries 01}
  (2019) 086} [\href{https://arxiv.org/abs/1808.02492}{{\ttfamily
  1808.02492}}].

\bibitem{Aharony:2018ics}
O.~Aharony, S.~Datta, A.~Giveon, Y.~Jiang and D.~Kutasov, \emph{{Modular
  covariance and uniqueness of $J\bar{T}$ deformed CFTs}},
  \href{https://doi.org/10.1007/JHEP01(2019)085}{\emph{JHEP} {\bfseries 01}
  (2019) 085} [\href{https://arxiv.org/abs/1808.08978}{{\ttfamily
  1808.08978}}].

\bibitem{Cardy:2015xaa}
J.~Cardy, \emph{{Quantum Quenches to a Critical Point in One Dimension: some
  further results}},
  \href{https://doi.org/10.1088/1742-5468/2016/02/023103}{\emph{J. Stat. Mech.}
  {\bfseries 1602} (2016) 023103}
  [\href{https://arxiv.org/abs/1507.07266}{{\ttfamily 1507.07266}}].

\bibitem{Giribet:2017imm}
G.~Giribet, \emph{{$T\bar{T}$-deformations, AdS/CFT and correlation
  functions}}, \href{https://doi.org/10.1007/JHEP02(2018)114}{\emph{JHEP}
  {\bfseries 02} (2018) 114}
  [\href{https://arxiv.org/abs/1711.02716}{{\ttfamily 1711.02716}}].

\bibitem{Kraus:2018xrn}
P.~Kraus, J.~Liu and D.~Marolf, \emph{{Cutoff AdS$_{3}$ versus the $
  T\overline{T} $ deformation}},
  \href{https://doi.org/10.1007/JHEP07(2018)027}{\emph{JHEP} {\bfseries 07}
  (2018) 027} [\href{https://arxiv.org/abs/1801.02714}{{\ttfamily
  1801.02714}}].

\bibitem{Aharony:2018vux}
O.~Aharony and T.~Vaknin, \emph{{The TT* deformation at large central charge}},
  \href{https://doi.org/10.1007/JHEP05(2018)166}{\emph{JHEP} {\bfseries 05}
  (2018) 166} [\href{https://arxiv.org/abs/1803.00100}{{\ttfamily
  1803.00100}}].

\bibitem{Guica:2019vnb}
M.~Guica, \emph{{On correlation functions in $J\bar T$-deformed CFTs}},
  \href{https://arxiv.org/abs/1902.01434}{{\ttfamily 1902.01434}}.

\bibitem{Shyam:2017znq}
V.~Shyam, \emph{{Background independent holographic dual to $T\bar{T}$ deformed
  CFT with large central charge in 2 dimensions}},
  \href{https://doi.org/10.1007/JHEP10(2017)108}{\emph{JHEP} {\bfseries 10}
  (2017) 108} [\href{https://arxiv.org/abs/1707.08118}{{\ttfamily
  1707.08118}}].

\bibitem{Cottrell:2018skz}
W.~Cottrell and A.~Hashimoto, \emph{{Comments on $T \bar T$ double trace
  deformations and boundary conditions}},
  \href{https://doi.org/10.1016/j.physletb.2018.09.068}{\emph{Phys. Lett.}
  {\bfseries B789} (2019) 251}
  [\href{https://arxiv.org/abs/1801.09708}{{\ttfamily 1801.09708}}].

\bibitem{Bzowski:2018pcy}
A.~Bzowski and M.~Guica, \emph{{The holographic interpretation of $J \bar
  T$-deformed CFTs}},
  \href{https://doi.org/10.1007/JHEP01(2019)198}{\emph{JHEP} {\bfseries 01}
  (2019) 198} [\href{https://arxiv.org/abs/1803.09753}{{\ttfamily
  1803.09753}}].

\bibitem{Taylor:2018xcy}
M.~Taylor, \emph{{TT deformations in general dimensions}},
  \href{https://arxiv.org/abs/1805.10287}{{\ttfamily 1805.10287}}.

\bibitem{Hartman:2018tkw}
T.~Hartman, J.~Kruthoff, E.~Shaghoulian and A.~Tajdini, \emph{{Holography at
  finite cutoff with a $T^2$ deformation}},
  \href{https://doi.org/10.1007/JHEP03(2019)004}{\emph{JHEP} {\bfseries 03}
  (2019) 004} [\href{https://arxiv.org/abs/1807.11401}{{\ttfamily
  1807.11401}}].

\bibitem{Shyam:2018sro}
V.~Shyam, \emph{{Finite Cutoff AdS$_{5}$ Holography and the Generalized
  Gradient Flow}}, \href{https://doi.org/10.1007/JHEP12(2018)086}{\emph{JHEP}
  {\bfseries 12} (2018) 086}
  [\href{https://arxiv.org/abs/1808.07760}{{\ttfamily 1808.07760}}].

\bibitem{Caputa:2019pam}
P.~Caputa, S.~Datta and V.~Shyam, \emph{{Sphere partition functions and cut-off
  AdS}},  \href{https://arxiv.org/abs/1902.10893}{{\ttfamily 1902.10893}}.

\bibitem{Gorbenko:2018oov}
V.~Gorbenko, E.~Silverstein and G.~Torroba, \emph{{dS/dS and $T\bar T$}},
  \href{https://arxiv.org/abs/1811.07965}{{\ttfamily 1811.07965}}.

\bibitem{Apolo:2018qpq}
L.~Apolo and W.~Song, \emph{{Strings on warped AdS$_{3}$ via $
  \mathrm{T}\bar{\mathrm{J}} $ deformations}},
  \href{https://doi.org/10.1007/JHEP10(2018)165}{\emph{JHEP} {\bfseries 10}
  (2018) 165} [\href{https://arxiv.org/abs/1806.10127}{{\ttfamily
  1806.10127}}].

\bibitem{Giveon:2017myj}
A.~Giveon, N.~Itzhaki and D.~Kutasov, \emph{{A solvable irrelevant deformation
  of AdS$_{3}$/CFT$_{2}$}},
  \href{https://doi.org/10.1007/JHEP12(2017)155}{\emph{JHEP} {\bfseries 12}
  (2017) 155} [\href{https://arxiv.org/abs/1707.05800}{{\ttfamily
  1707.05800}}].

\bibitem{Asrat:2017tzd}
M.~Asrat, A.~Giveon, N.~Itzhaki and D.~Kutasov, \emph{{Holography Beyond AdS}},
  \href{https://doi.org/10.1016/j.nuclphysb.2018.05.005}{\emph{Nucl. Phys.}
  {\bfseries B932} (2018) 241}
  [\href{https://arxiv.org/abs/1711.02690}{{\ttfamily 1711.02690}}].

\bibitem{Baggio:2018gct}
M.~Baggio and A.~Sfondrini, \emph{{Strings on NS-NS Backgrounds as Integrable
  Deformations}}, \href{https://doi.org/10.1103/PhysRevD.98.021902}{\emph{Phys.
  Rev.} {\bfseries D98} (2018) 021902}
  [\href{https://arxiv.org/abs/1804.01998}{{\ttfamily 1804.01998}}].

\bibitem{Babaro:2018cmq}
J.~P. Babaro, V.~F. Foit, G.~Giribet and M.~Leoni, \emph{{$ T\overline{T} $
  type deformation in the presence of a boundary}},
  \href{https://doi.org/10.1007/JHEP08(2018)096}{\emph{JHEP} {\bfseries 08}
  (2018) 096} [\href{https://arxiv.org/abs/1806.10713}{{\ttfamily
  1806.10713}}].

\bibitem{Chakraborty:2018aji}
S.~Chakraborty, \emph{{Wilson loop in a $T\bar{T}$ like deformed
  $\rm{CFT}_2$}},
  \href{https://doi.org/10.1016/j.nuclphysb.2018.12.003}{\emph{Nucl. Phys.}
  {\bfseries B938} (2019) 605}
  [\href{https://arxiv.org/abs/1809.01915}{{\ttfamily 1809.01915}}].

\bibitem{Araujo:2018rho}
T.~Araujo, E.~{\'O}~Colg{\'a}in, Y.~Sakatani, M.~M. Sheikh-Jabbari and
  H.~Yavartanoo, \emph{{Holographic integration of $T \bar{T}$ \& $J \bar{T}$
  via $O(d,d)$}},  \href{https://arxiv.org/abs/1811.03050}{{\ttfamily
  1811.03050}}.

\bibitem{Chakraborty:2018kpr}
S.~Chakraborty, A.~Giveon, N.~Itzhaki and D.~Kutasov, \emph{{Entanglement
  beyond AdS}},
  \href{https://doi.org/10.1016/j.nuclphysb.2018.08.011}{\emph{Nucl. Phys.}
  {\bfseries B935} (2018) 290}
  [\href{https://arxiv.org/abs/1805.06286}{{\ttfamily 1805.06286}}].

\bibitem{Donnelly:2018bef}
W.~Donnelly and V.~Shyam, \emph{{Entanglement entropy and $T \overline{T}$
  deformation}},
  \href{https://doi.org/10.1103/PhysRevLett.121.131602}{\emph{Phys. Rev. Lett.}
  {\bfseries 121} (2018) 131602}
  [\href{https://arxiv.org/abs/1806.07444}{{\ttfamily 1806.07444}}].

\bibitem{Baggio:2018rpv}
M.~Baggio, A.~Sfondrini, G.~Tartaglino-Mazzucchelli and H.~Walsh, \emph{{On
  $T\bar{T}$ deformations and supersymmetry}},
  \href{https://arxiv.org/abs/1811.00533}{{\ttfamily 1811.00533}}.

\bibitem{Chang:2018dge}
C.-K. Chang, C.~Ferko and S.~Sethi, \emph{{Supersymmetry and $T \overline{T}$
  Deformations}},  \href{https://arxiv.org/abs/1811.01895}{{\ttfamily
  1811.01895}}.

\bibitem{Santilli:2018xux}
L.~Santilli and M.~Tierz, \emph{{Large N phase transition in $ T\overline{T} $
  -deformed 2d Yang-Mills theory on the sphere}},
  \href{https://doi.org/10.1007/JHEP01(2019)054}{\emph{JHEP} {\bfseries 01}
  (2019) 054} [\href{https://arxiv.org/abs/1810.05404}{{\ttfamily
  1810.05404}}].

\bibitem{Bonelli:2018kik}
G.~Bonelli, N.~Doroud and M.~Zhu, \emph{{$T \bar{T}$-deformations in closed
  form}}, \href{https://doi.org/10.1007/JHEP06(2018)149}{\emph{JHEP} {\bfseries
  06} (2018) 149} [\href{https://arxiv.org/abs/1804.10967}{{\ttfamily
  1804.10967}}].

\bibitem{Conti:2018jho}
R.~Conti, L.~Iannella, S.~Negro and R.~Tateo, \emph{{Generalised Born-Infeld
  models, Lax operators and the $ \mathrm{T}\overline{\mathrm{T}} $
  perturbation}}, \href{https://doi.org/10.1007/JHEP11(2018)007}{\emph{JHEP}
  {\bfseries 11} (2018) 007}
  [\href{https://arxiv.org/abs/1806.11515}{{\ttfamily 1806.11515}}].

\bibitem{Chen:2018eqk}
B.~Chen, L.~Chen and P.-X. Hao, \emph{{Entanglement entropy in
  $T\overline{T}$-deformed CFT}},
  \href{https://doi.org/10.1103/PhysRevD.98.086025}{\emph{Phys. Rev.}
  {\bfseries D98} (2018) 086025}
  [\href{https://arxiv.org/abs/1807.08293}{{\ttfamily 1807.08293}}].

\bibitem{Conti:2018tca}
R.~Conti, S.~Negro and R.~Tateo, \emph{{The $ \mathrm{T}\overline{\mathrm{T}} $
  perturbation and its geometric interpretation}},
  \href{https://doi.org/10.1007/JHEP02(2019)085}{\emph{JHEP} {\bfseries 02}
  (2019) 085} [\href{https://arxiv.org/abs/1809.09593}{{\ttfamily
  1809.09593}}].

\bibitem{ongoing}
S.~Chakraborty, S.~Datta, A.~Giveon, Y.~Jiang and D.~Kutasov, \emph{work in
  progress}, .

\bibitem{note}
A.~Giveon, \emph{{Comments on $T\bar T$, $J\bar T$ and String Theory}},
  {\emph{{appears in a coordinated submission to the arXiv}} }.

\bibitem{Dong:2014tsa}
X.~Dong, D.~Z. Freedman and Y.~Zhao, \emph{{Explicitly Broken Supersymmetry
  with Exactly Massless Moduli}},
  \href{https://doi.org/10.1007/JHEP06(2016)090}{\emph{JHEP} {\bfseries 06}
  (2016) 090} [\href{https://arxiv.org/abs/1410.2257}{{\ttfamily 1410.2257}}].

\end{thebibliography}\endgroup

\end{document}